\documentclass[12pt,preprint]{aastex}

\shorttitle{Quark deconfinement and implications}
\shortauthors{Bombaci et al.}

\def\be{\begin{equation}}
\def\ee{\end{equation}}
\def\bq{\begin{eqnarray}}
\def\eq{\end{eqnarray}}
\def\R{\mbox{${\cal R}$}}

\def\exo07{\mbox{EXO 0748-676}}
\def\1e12{\mbox{1E 1207.4-5209 }}
\def\rx18{\mbox{RX J1856.5-3754}}
\def\etal{\mbox{\it et al.}}
\begin{document}


\title{Quark deconfinement and implications for the radius \\ 
           and  the limiting mass of compact stars} 


\author{Ignazio Bombaci\altaffilmark{1},  Irene Parenti\altaffilmark{1}, 
and Isaac Vida\~na\altaffilmark{1}}


\altaffiltext{1}{Dipartimento di Fisica ``Enrico Fermi'', Universit\'a di Pisa,
and INFN Sezione Pisa, via Buonarroti 2, I-56127 Pisa, Italy.}


\begin{abstract}
We study the consequences  of the hadron-quark deconfinement phase transition 
in stellar compact objects when finite size effects between the deconfined quark 
phase and the hadronic phase are taken into account.  
We show that above a threshold value of the central pressure  (gravitational mass)  
a neutron star is metastable to the decay (conversion) to a hybrid neutron star 
or to a strange star.  The {\it mean-life time}  of the metastable configuration  
dramatically depends on the value of the stellar central pressure.    
We explore the consequences of the metastability of ``massive'' neutron stars 
and of the existence of stable compact  quark  stars (hybrid neutron stars or 
strange stars) on the concept of limiting  mass of compact stars. 
We discuss the implications of our scenario on the interpretation 
of the stellar mass and radius extracted from the spectra of several 
X-ray compact sources.
%%%%%%%%%% REV>VERSION 01/04  %%%%%%%%%
Finally, we show that our scenario implies, as a natural consequence 
a two step-process which is able to explain the inferred ``delayed'' connection 
between supernova explosions  and GRBs, giving also the correct 
energy to power GRBs.   
\end{abstract}


%% subject header.  Objects should be in the appropriate "individual"
%% headers (e.g. quasars: individual, stars: individual, etc.) with the
%% additional provision that the total number of headers, including each
%% individual object, not exceed six.  The \objectname{} macro, and its
%% alias \object{}, is used to mark each object.  The macro takes the object
%% name as its primary argument.  This name will appear in the paper
%% and serve as the link's anchor in the electronic edition if the name
%% is recognized by the data centers.  The macro also takes an optional
%% argument in parentheses in cases where the data center identification
%% differs from what is to be printed in the paper.



\section{Introduction} 

One of the most fascinating enigma in modern astrophysics concerns  the 
true nature of the ultra-dense compact objects called {\it neutron stars} (NS).  
The bulk properties and the internal structure of these stars 
chiefly depends upon the equation of state (EOS) of dense hadronic matter. 
Different models for the EOS of dense matter  predict  a neutron star  
maximum mass ($M_{max}$) in the range of   1.4 -- 2.2 $M_\odot$, and a  
corresponding central density  in range of 4 -- 8 times the saturation density  
($\rho_0 \sim 2.8 \times 10^{14}$g/cm$^3$)  of nuclear matter  
(e.g.  Shapiro \& Teukolsky 1983; Haensel 2003).   
In the case  of a star with $ M \sim 1.4~M_\odot$, different EOS models predict 
a radius  in the range of 7 -- 16 km   (Shapiro \& Teukolsky 1983; Haensel 2003; 
Dey et al. 1998).      

In a simplistic and conservative picture  the core of a neutron star  is modeled 
as a uniform fluid of neutron rich nuclear matter in equilibrium with respect 
to the weak interaction ($\beta$-stable nuclear matter).  
However, due to the large value of the stellar central density and to the 
rapid increase of the nucleon chemical potentials with density,   
hyperons ($\Lambda$, $\Sigma^{-}$, $\Sigma^{0}$, $\Sigma^{+}$, $\Xi^{-}$ 
and $\Xi^{0}$ particles) are expected to appear in the inner core of the star.    
Other {\it exotic} phases of hadronic matter such as  a Bose-Einstein condensate 
of negative pion ($\pi^-$) or negative kaon ($K^-$)  could be present in 
the inner part of the star.  

  According to Quantum Chromodynamics (QCD)  a phase transition from hadronic 
matter to a deconfined quark phase  should  occur at a density of a few times nuclear 
matter saturation density.  Consequently, the core of the more massive neutron stars  
is one of the best candidates in the Universe where such deconfined phase of quark 
matter (QM) could be found. This possibility was realized by several researchers soon 
after the introduction of quarks as the fundamentals building 
blocks of hadrons (Ivanenko \& Kurdgelaidze 1969; Itoh 1970; Iachello et al. 1974;
Collins \& Perry 1975; Baym \& Chin 1976; Keister \& Kisslinger 1976). 
Since $\beta$-stable hadronic matter posses two conserved ``charges'' ({\it i.e.,} 
electric charge and baryon number) the quark-deconfinement phase transition 
proceeds through a mixed phase over a finite range of pressures and densities 
according to the Gibbs' criterion for phase equilibrium  
(Glendenning 1992; M\"{u}ller \& Serot 1995).   
At the onset of the mixed phase,  quark matter droplets form a Coulomb lattice 
embedded in a sea of hadrons and in a roughly uniform sea of electrons and muons. 
As the pressure increases various geometrical shapes (rods, plates) of the less abundant 
phase immersed in the dominant one are expected. Finally the system turns into 
uniform quark matter  at the highest pressure of the mixed phase 
(Heiselberg et al. 1993; Voskresensky et al. 2003).         
Compact stars  which possess a  ``quark matter core'' either as a 
mixed phase of deconfined quarks and hadrons or as a pure quark matter phase 
are called {\it Hybrid Neutron Stars} or shortly {\it Hybrid Stars} (HyS)  
(Glendenning 1996; Drago \& Lavagno 2001).           
In the following of this paper, the  more {\it conventional} neutron stars  
in which no fraction of quark matter is present, will be referred to as  
{\it pure Hadronic Stars} (HS).  
 
Even more intriguing than the existence of a quark core in a neutron star, 
is the possible existence of a new family of compact stars consisting completely 
of a deconfined mixture of {\it up} ({\it u}),  {\it down} ({\it d})  and  {\it strange } ({\it s}) 
quarks  (together with an appropriate number of electrons to guarantee electrical 
neutrality)  satisfying the Bodmer--Witten hypothesis  
(Bodmer 1971; Witten 1984; see also Terazawa 1979).      
Such compact stars have been called {\it strange quark stars} or shortly 
{\it strange stars} (SS) (Alcock et al. 1986; Haensel et al. 1986)  
and their constituent matter as {\it strange quark matter} (SQM) 
(Farhi \& Jaffe 1984; Madsen 1999).  
Presently there is no unambiguous proof about the existence of 
strange stars, however, a  sizable amount of observational data collected by 
the new generations of X-ray satellites, is providing a growing body of evidence 
for their  possible existence  (Bombaci 1997; Cheng et al. 1998; Li et al. 1999a;
Li et al. 1999b; Xu 2002;  Drake et al. 2002).    
It is generally believed that the unambiguous 
identification of a strange star will imply that all pulsars must be strange stars. 
In the present work we argue that the possible existence of strange stars does not 
conflict with the existence of  {\it conventional} neutron stars (pure Hadronic Stars).   

Present accurate determinations of compact star masses in radio pulsar 
binaries (Thorsett \& Chakrabarty 1999)  permit to rule out only 
{\it extremely soft} EOS, {\it i.e.} those giving $M_{max}$ less 
than about 1.45~$M_\odot$.      
However, in at least two  accreting X-ray binaries it has been found evidence 
for compact stars with higher  masses. The first of these star is Vela X-1, 
with a reported mass $1.88 \pm 0.13 M_\odot$ (Quaintrell et al.  2003), 
the second is  Cygnus X-2, with a reported mass of $1.78 \pm 0.23 M_\odot$  
(Orosz \& Kuulkers 1999). 
Unfortunately, mass determinations in X-ray binaries are affected by large 
uncertainties (van Kerkwijk et al. 1995), therefore the previous quoted 
``high mass values'' should always be handled with care.   
In addition to mass determination, existing observational data on the spin 
frequency of millisecond pulsars and on the thermal evolution of neutron stars 
do not put severe constraints  on  the EOS of dense matter.  
Fortunately, this situation is improving in the last few years. 
In fact, the extraordinary spectroscopic capabilities of the instruments on board 
Chandra X-ray and XMM-Newton satellites, are giving the unique possibility to perform 
accurate measurements of the gravitational red-shift in the spectral lines of 
a few compact X-ray sources. This provide informations on the mass to radius ratio for 
compact stars and will help us to understand the true nature of these compact objects.   

In this work we study the effects of the hadron-quark deconfinement phase transition 
in stellar compact objects.    
We show that when finite size effects at the interface between the quark- and 
the hadron-phase are taken into account, pure Hadronic Stars, above a threshold  
value of the central pressure (gravitational mass),   are  metastable 
to the {\it decay}  ({\it conversion})  to  hybrid neutron stars or to  strange stars 
(depending on the properties of EOS for  quark matter).  
The {\it mean-life time} of the metastable stellar configuration is related to the 
quantum nucleation time to form a drop of quark matter in the stellar center, 
and  dramatically depends on the value of the stellar central pressure.   
We explore the consequences of the metastability of ``massive'' pure Hadronic Stars  
and the existence of stable compact quark  stars  (hybrid neutron stars or 
strange stars)  on the concept of limiting mass of compact stars.  
Next,  we discuss the implications of our scenario in the interpretation of the 
mass-radius  constraints extracted from the spectra of several X-ray compact sources. 
Finally, we discuss the implications of our scenario for Gamma Ray Bursts.

\section{Quantum nucleation of quark matter in hadronic stars}  

%%%  REV.VERSION 01/04 %%%%%%%%
Nucleation of quark matter in neutron stars has been studied by many authors. 
Most of the earlier studies on this subject (Horvath et al. 1992; Horvath 1994; 
Olesen \& Madsen 1994;  Heiselberg 1995; Grassi 1998)  have been restricted to the 
case of  thermal nucleation in hot and dense hadronic matter. 
In these studies, it was found that the prompt formation of a critical size drop 
of quark matter  via thermal activation is possible above a temperature 
of about 2 -- 3 MeV. 
As a consequence,  it was inferred  that pure hadronic stars are converted 
to strange stars or to hybrid stars within the first seconds after their birth.   
It was also suggested that the large amount of energy liberated in this process 
could play a crucial role in type-II supernova explosions (Benvenuto \& Horvath 1989).     

All the studies on quark matter nucleation mentioned above, have neglected an  
important physical aspect which charecterizes dense matter in a newly born  
{\it neutron star}\,: the trapping of neutrinos  in the stellar core.  
Neutrino trapping has a sizeable influence on the {\it stiffness} of the EOS  and, 
consequently, on the structural properties of the protoneutron star 
(Bombaci 1996;  Prakash et al. 1997).      
In particular, it has been found that neutrino trapping significantly shifts 
the critical baryon density for the quark deconfinement phase transition 
to higher values with respect to the neutrino-free case (Prakash et al. 1997; 
Lugones \& Benvenuto 1998).  
In addition, neutrino trapping decreases the value of the central density 
of the stellar maximum mass configuration with respect to the neutrino-free 
case (Prakash et al. 1997).     
Thus, the formation of a metastable supercompressed phase of hadronic matter 
is strongly inhibited  in a protoneutron star.    
A detailed study of quark matter nucleation in hot and dense hadronic matter 
with trapped neutrinos will be presented in a forthcoming work (Bombaci et al. 2004).  

In the present paper, we assume that the compact star survives the early stages 
of its evolution as a pure hadronic star, and we study quark matter nucleation 
in cold (T = 0) neutrino-free hadronic matter.     
%%%%%%%%%%%%%%%%%%%%%%%%%%%%%%%%%%%%%%%%%%%%

In  bulk matter the quark-hadron mixed phase begins at the {\it static transition  
point} defined according to the Gibbs' criterion for phase equilibrium 
\begin{equation}
\mu_H = \mu_Q \equiv \mu_0 \, , ~~~~~~~~~~~~
P_H(\mu_0) = P_Q(\mu_0) \equiv P_0  \,     
\label{eq:eq1}
\end{equation}
where 
\begin{equation}
  \mu_H = \frac{\varepsilon_H + P_H}{n_{b,H}}  \, , ~~~~~~~~~~~~
  \mu_Q = \frac{\varepsilon_Q + P_Q}{n_{b,Q}}  
\label{eq:eq2}
\end {equation} 
are the chemical potentials for the hadron and quark phase respectively, 
$\varepsilon_H$ ($\varepsilon_Q$),  $P_H$ ($P_Q$)  and $n_{b,H}$  ($ n_{b,Q}$)
denote respectively the total ({\it i.e.,}  including leptonic contributions) energy 
density,  the total pressure and baryon number density  for the hadron (quark)  
phase,  in the case of cold  matter.   

Let us now consider the more realistic situation in which one takes into account 
the energy cost due to finite size effects  in creating a drop of deconfined quark matter 
in the hadronic environment.  As a consequence of these effects,  the formation 
of a critical-size drop of QM is not immediate and it is necessary to have 
an overpressure  $\Delta P =  P - P_0$  with respect to the static the transition point.  
Thus,  above  $P_0$, hadronic matter is  in a metastable state,  and the formation 
of a real drop of quark matter occurs via a quantum nucleation mechanism.  
A sub-critical (virtual) droplet of deconfined quark matter moves back and 
forth in the potential energy well separating the two matter phases 
(see discussion below) on a time scale 
$\nu_0^{-1} \sim 10^{-23}$ seconds, which is set by the strong interactions. 
This time scale is many orders of magnitude shorter than the typical time scale 
for the weak interactions,  therefore 
quark flavor must be conserved during the deconfinement transition. 
We will refer to this form of deconfined matter,  in which the 
flavor content is equal to that of the $\beta$-stable hadronic system 
at the same pressure,  as the  Q*-phase. 
Soon afterwards a critical size drop of quark matter is formed the weak interactions 
will have enough time to act, changing the quark flavor fraction of the deconfined 
droplet to lower its energy, and a droplet of $\beta$-stable SQM is formed 
(hereafter the Q-phase).  For example, if quark deconfinement occurs in 
$\beta$-stable  nuclear matter (non-strange hadronic matter), it will produce 
a two-flavor ($u$ and $d$ ) quark matter droplet having 
\begin{equation}
n_u / n_d = (1 + x_p)/(2-x_p) \,  ,     
\label{eq:eq3}
\end{equation}
$n_u$ and $n_d$ being the {\it up} and {\it down} quark number densities respectively, 
and $x_p$ the proton fraction in the $\beta$-stable hadronic phase.   
In the more general case in which the hadronic phase has a strangeness content 
({\it e.g.,} hyperonic matter), the deconfinement transition will form a droplet 
of strange matter with a flavor content equal to that of the $\beta$-stable 
hadronic system at the same pressure,  according to the relation:
\begin{equation}
\left( \begin{array}{c}
x_u \\ x_d \\ x_s
\end{array} \right)
=
\left( \begin{array}{cccccccc}
2 & 1 & 1 & 2 & 1 & 0 & 1 & 0 \\
1 & 2 & 1 & 0 & 1 & 2 & 0 & 1 \\
0 & 0 & 1 & 1 & 1 & 1 & 2 & 2
\end{array} \right)
\left( \begin{array}{c}
x_p \\ x_n \\ x_{\Lambda} \\ x_{\Sigma^+} \\
x_{\Sigma^0} \\ x_{\Sigma^-} \\ x_{\Xi^0} \\ x_{\Xi^-}
\end{array} \right) \, ,
\label{eq:eq4}
\end{equation}
where  $x_i = n_i/n_b$ are the concentrations of the different particle species.  

In the present work, we have adopted rather common models for describing both 
the hadronic and the quark phase of dense matter. For the hadronic phase we used 
models which are based on a relativistic 
lagrangian of hadrons interacting via the exchange of sigma, rho 
and omega mesons. The parameters adopted are the standard ones 
(Glendenning \& Moszkowski 1991). 
Hereafter we refer to this model as the GM equation of state (EOS).    
For the quark phase we have adopted a phenomenological EOS (Farhi \& Jaffe 1984) which is 
based on the MIT bag model for hadrons. 
The parameters here are: the mass $m_s$ of the strange quark, the so-called 
pressure of the vacuum $B$ (bag constant) and the QCD structure constant  $\alpha_s$.  
For all the quark matter model used in the present work, we take  $m_u = m_d =0$, 
$m_s = 150$~MeV  and $\alpha_s = 0$.   

\begin{figure}
%%\resizebox{\hsize}{!}{\includegraphics{figure1.eps}}
%%%\plotone{figure1.eps} 
\plotone{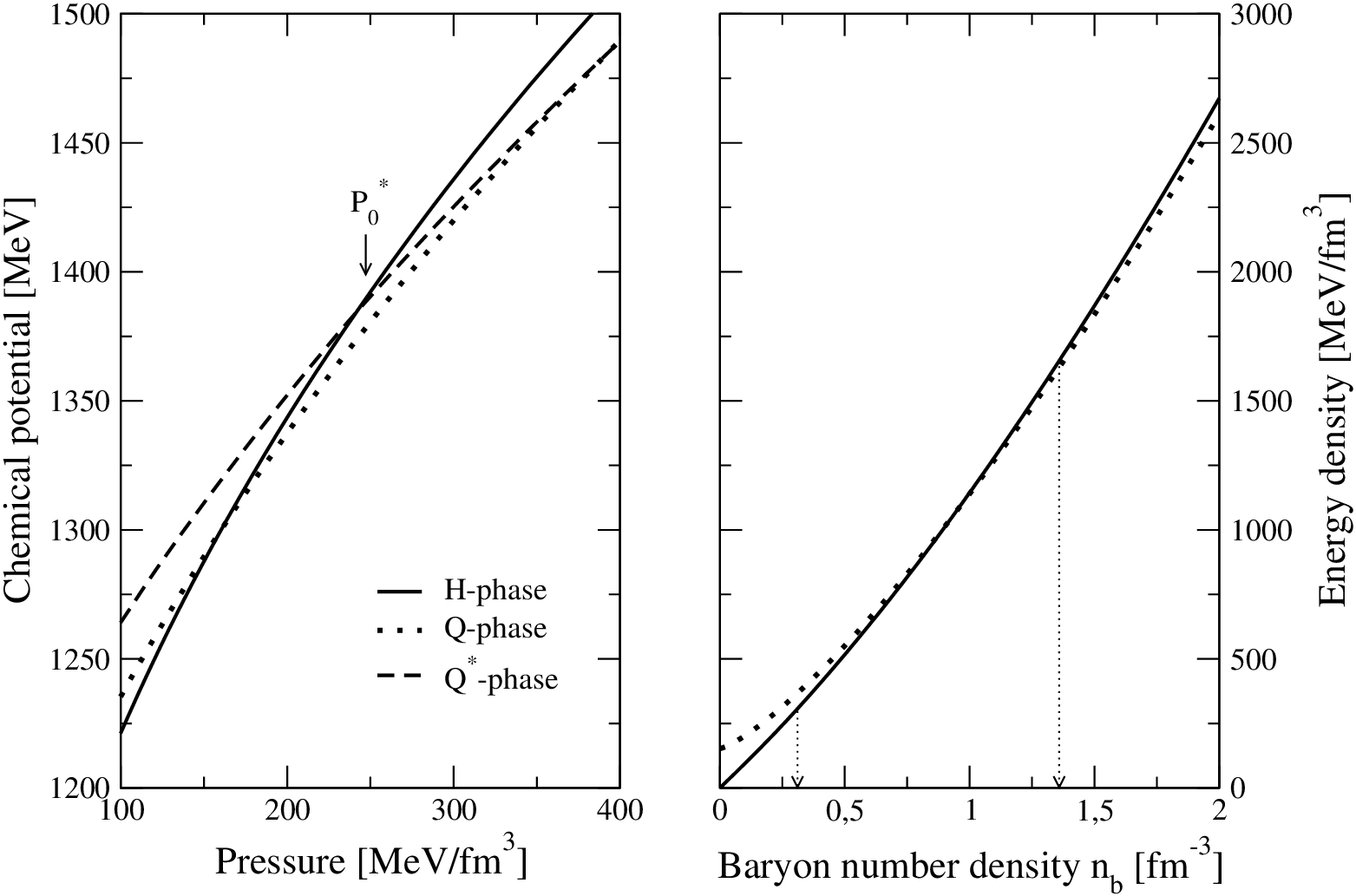}  
\caption{Chemical potentials of the three phases of matter (H, Q, and Q$^*$), 
as defined by Eq. (2) as a function of the total pressure (left panel); and energy 
density of the H- and Q-phase  as a function of the baryon number density (right panel). 
The hadronic phase is described with the GM3 model whereas for the 
Q and Q$^*$ phases is employed the MIT-like bag model with 
$m_s=150$ MeV, $B=152.45$ MeV/fm$^3$ and $\alpha_s=0$. 
The vertical lines arrows on the right panel indicate the beginning and the end of 
the mixed hadron-quark phase defined according to the Gibbs criterion for phase 
equilibrium. On the left panel $P_0^*$ denotes the static transition point.}
\label{fig:fig1}
\end{figure}

In the left panel of  Fig.\ \ref{fig:fig1},  we show the chemical potentials,  defined 
according to Eq.\ (\ref{eq:eq2}), as a function of the total pressure for the three phases 
of matter (H, Q*, and Q) discussed above. In the right panel of the same figure, 
we plot the  energy densities for the H- and Q-phase  as a function of the 
corresponding baryon number densities. 
Both panels in Fig.\ \ref{fig:fig1} are relative to the GM3 model for the  EOS for 
the H-phase and to the MIT bag model EOS for the Q and Q* phases 
with $B=152.45$ MeV/fm$^{3}$.      
 
To calculate the nucleation rate of quark matter in the hadronic medium  
we use the Lifshitz--Kagan quantum nucleation theory (Lifshitz \& Kagan 1972) 
in the relativistic form given by Iida \& Sato (1997).    
The QM droplet is supposed to be a sphere of radius $\cal R$ and  
its quantum fluctuations are described by the lagrangian  
\begin{equation}
 L({\cal R},{\dot {\cal R}})  = - {\cal M}({\cal R})  c^2 \sqrt{1 - ({\dot {\cal R}} /c)^2} 
                                               + {\cal M}({\cal R}) c^2 - U({\cal R}) \, ,
\label{eq:eq5}
\end{equation}
where $ {\cal M}({\cal R}) $ is the effective mass of the QM droplet, and $U({\cal R}) $
its  potential energy. 
%%%  REV.VERSION 01/04 %%%%%%%%
Within the Lifshitz--Kagan quantum nucleation theory, one assumes that 
the phase boundary  ({\it i.e.} the droplet surface) moves slowly compared to 
the high sound velocity of the medium ($\dot{\R} << v_s \sim c$). 
Thus the number density of each phase adjust adiabatically to the fluctuations 
of the droplet radius, and the system retains pressure equilibrium between the 
two phases. 
Thus, the droplet effective mass is given by (Lifshitz \& Kagan 1972; Iida \& Sato 1997)  
\begin{equation}
{\cal M}(\R) = 4 \pi \rho_H \Big(1 - {n_{b,Q*}\over n_{b,H}}\Big)^2  \R^3  \, , 
\label{eq:eq6}
\end{equation}
$\rho_H$ being the hadronic mass density,  
$n_{b,H}$ and $n_{b,Q*}$ are the baryonic number densities at a same pressure 
in the hadronic and  Q*-phase,  respectively. 
The potential energy is given by  (Lifshitz \& Kagan 1972; Iida \& Sato 1997)    
\begin{equation}
      U(\R)   =    {4 \over 3} \pi  \R^3 n_{b,Q*} (\mu_{Q*}  - \mu_H)   + 
                           4 \pi \sigma \R^2   +  8 \pi \gamma \R\, ,
\label{eq:eq7}
\end{equation}
where  $\mu_H$ and $\mu_{Q*}$ are 
the hadronic and quark chemical potentials at a fixed pressure $P$
and $\sigma$ is the surface tension for the surface separating the 
quark phase  from the hadronic phase.  
%%%  REV.VERSION 01/04 %%%%%%%%
The value of the surface tension $\sigma$ is poorly known, 
and typical values used in the literature range within
10--50 MeV/fm$^2$ (Heiselberg et al. 1993; Iida \& Sato 1997).  
The third term in  Eq.\ (\ref{eq:eq7}),  $E_{curv} = 8 \pi \gamma \R$, 
is the so called {\it curvature energy}.  
It has been shown that $E_{curv}$ plays an important role in the thermal 
nucleation process of quark matter in hot hadronic matter (Horvath 1994; 
Olesen \& Madsen 1994).   Specifically, the curvature term increases the minimum 
temperature for thermal nucleation with respect to the case 
$\gamma = 0$ (Horvath 1994). 
Most of the results  reported in the present study have been obtained taking 
$\gamma = 0$ in Eq.\ (\ref{eq:eq7}). In a few cases (see Tab. 3), we considered 
$\gamma \neq 0$ to investigate the influence of the curvature energy on 
quantum nucleation in cold hadronic matter.  

In the previous expression (\ref{eq:eq7}) for the droplet potential 
energy, we neglected the terms connected with the electrostatic  energy.  
Detailed calculations by Iida \& Sato (1997)  have demonstrated that 
the contribution of these  terms to Eq.\ (\ref{eq:eq7})   
can be safely omitted since the screening action by leptons 
nearly compensate the effect of the positive charged droplet.    
%%%%%%%%%%%%%%%%%%%%%%%%%%%%%%%%

The process of formation of a bubble having a critical radius, 
can be computed using a semiclassical approximation. The procedure
is rather straightforward. First one computes, 
using the well known Wentzel--Kramers--Brillouin (WKB) approximation, 
the ground state energy $E_0$ and the oscillation frequency $\nu_0$ 
of the virtual QM drop in the potential well $U(\R)$. 
Then it is possible to calculate in a relativistic framework 
the probability of tunneling as (Iida \& Sato 1997)
\begin{equation}
p_0=\exp \Big[-{A(E_0)\over \hbar}\Big]
\label{eq:eq8}
\end{equation}
where $A$ is the action under the potential barrier 

\begin{equation}
A(E)  = {2\over c}\int_{\R_-}^{\R_+} \left\{ [2{\cal M}(\R)c^2 + E -U(\R)] 
            \times  [U(\R)-E] \right\}^{1/2} d\R   \, ,   
\label{eq:eq9}
\end{equation}
$\R_\pm$ being  the classical turning points. 

The nucleation time is then equal to 
\begin{equation}
\tau = (\nu_0 p_0 N_c)^{-1}\, ,
\label{eq:eq10}
\end{equation}
where $N_c$ is the number of virtual centers of droplet formation 
in the innermost region of the star.  
Following the simple estimate given in Iida \& Sato (1997), we take  $N_c = 10^{48}$.  
The uncertainty in the value of $N_c$ is expected to be within one or two orders 
of magnitude.  
In any case, all the qualitative features of our scenario will be not affected by the 
uncertainty in the value of $N_c$.

\section{Results}

\begin{figure}
%%\resizebox{\hsize}{!}{\includegraphics{figure2.eps}}
%%\plotone{figure2.eps} 
\plotone{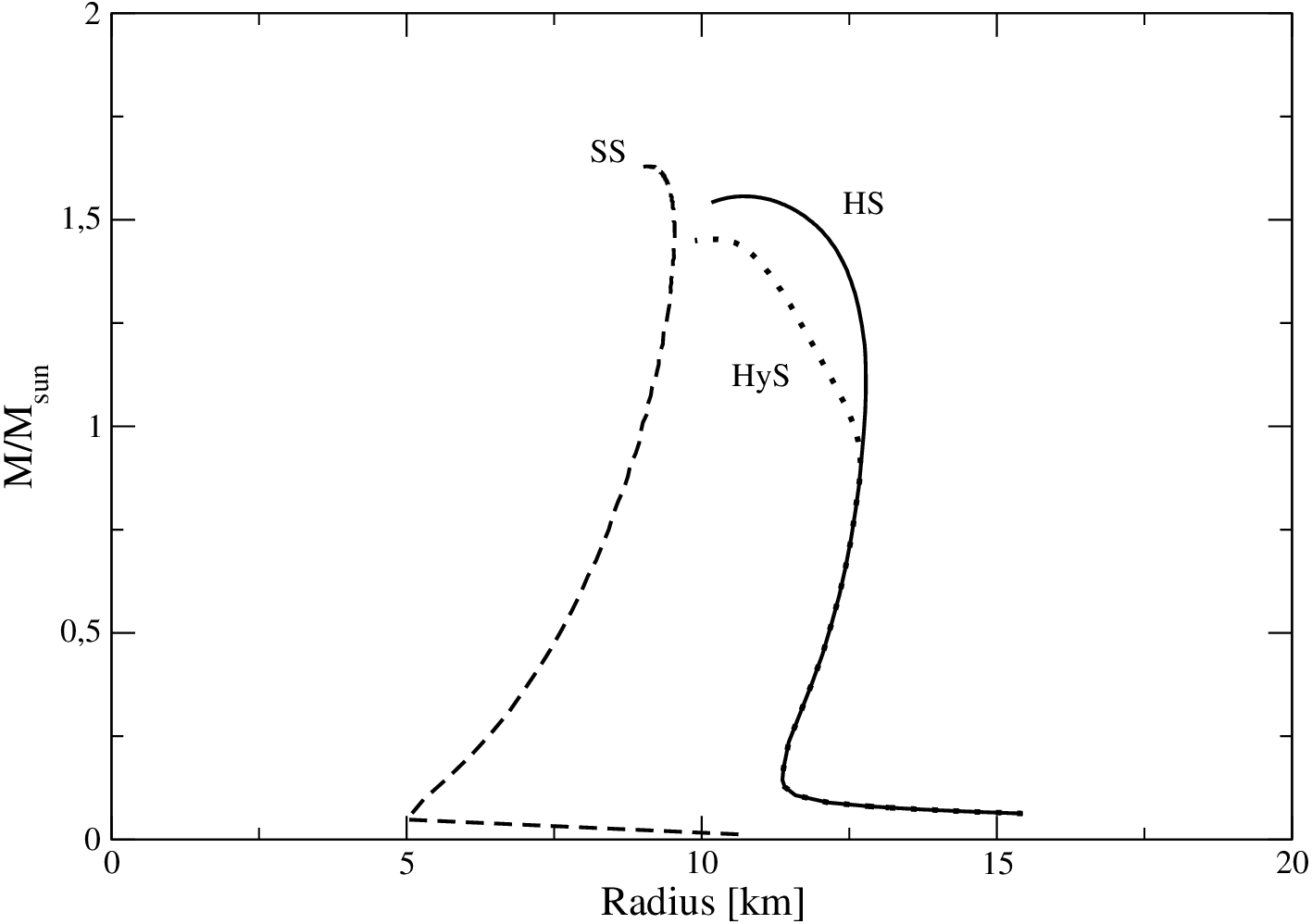}  
\caption{Mass-radius relations for the three types of compact objects discussed 
in the text: 
Hadronic Star (HS), Hybrid Star (HyS) and Strange Star (SS). 
The hadronic phase is described with the GM3 model while the pure quark phase is 
described by the MIT-like bag model with $m_u=m_d=0$, $m_s=150$ MeV, $\alpha_s=0$ 
and  $B=136.62 (69.47)$ MeV/fm$^3$ for the hybrid star (strange star).}
\label{fig:fig2}
\end{figure}

To begin with we show in Fig.\ \ref{fig:fig2}  the typical mass-radius (MR) relations 
for the three possible types of compact stars discussed before.  
The curve labeled with HS represents the MR relation 
for  pure hadronic stars containing an hyperonic core obtained with the 
GM3 model for the EOS of dense matter.  
The curve labeled HyS depicts the MR relation for  hybrid neutron stars  
where the hadronic phase is described by the same GM3 model for the EOS and the 
quark phase by the MIT-bag like model with  
%%%%%%%$m_u=m_d = 0$, $m_s = 150$ MeV,  $\alpha_s = 0$ and  
$B = 136.62$ MeV/fm$^3$. 
Finally, if we assume, for example,  $B = 69.47$ MeV/fm$^3$  (with the remaining 
parameters for quark phase unchanged with respect to the previous case), 
SQM fulfils  the Bodmer-Witten hypothesis  and 
one has the strange star sequence depicted by the curve SS in Fig.\ \ref{fig:fig2}.   
As it appears,  stars having a deconfined quark content (HyS or SS)  are more 
compact than purely hadronic stars (HS).  

 In our scenario, we consider a purely hadronic star whose central pressure  
is increasing due to spin-down or due to mass accretion, 
{\it e.g.,} from the material left by the supernova explosion (fallback disc), 
from a companion star or from the interstellar medium.    
As the central pressure  exceeds  the threshold value $P_0^*$ at the  static 
transition point,  a virtual drop of quark matter in the Q*-phase can be formed 
in the central region of the star.   As soon as a real drop of Q*-matter is formed, 
it will grow very rapidly 
and the original Hadronic Star will be converted to 
and Hybrid Star or to a Strange Star, depending on the detail of the EOS 
for quark matter employed to model the phase transition (particularly depending 
on the value of the parameter $B$ within  the model adopted in the present study).  

\begin{figure}
%%\resizebox{\hsize}{!}{\includegraphics{figure3.eps}}
%%%\plotone{figure3.eps} 
\plotone{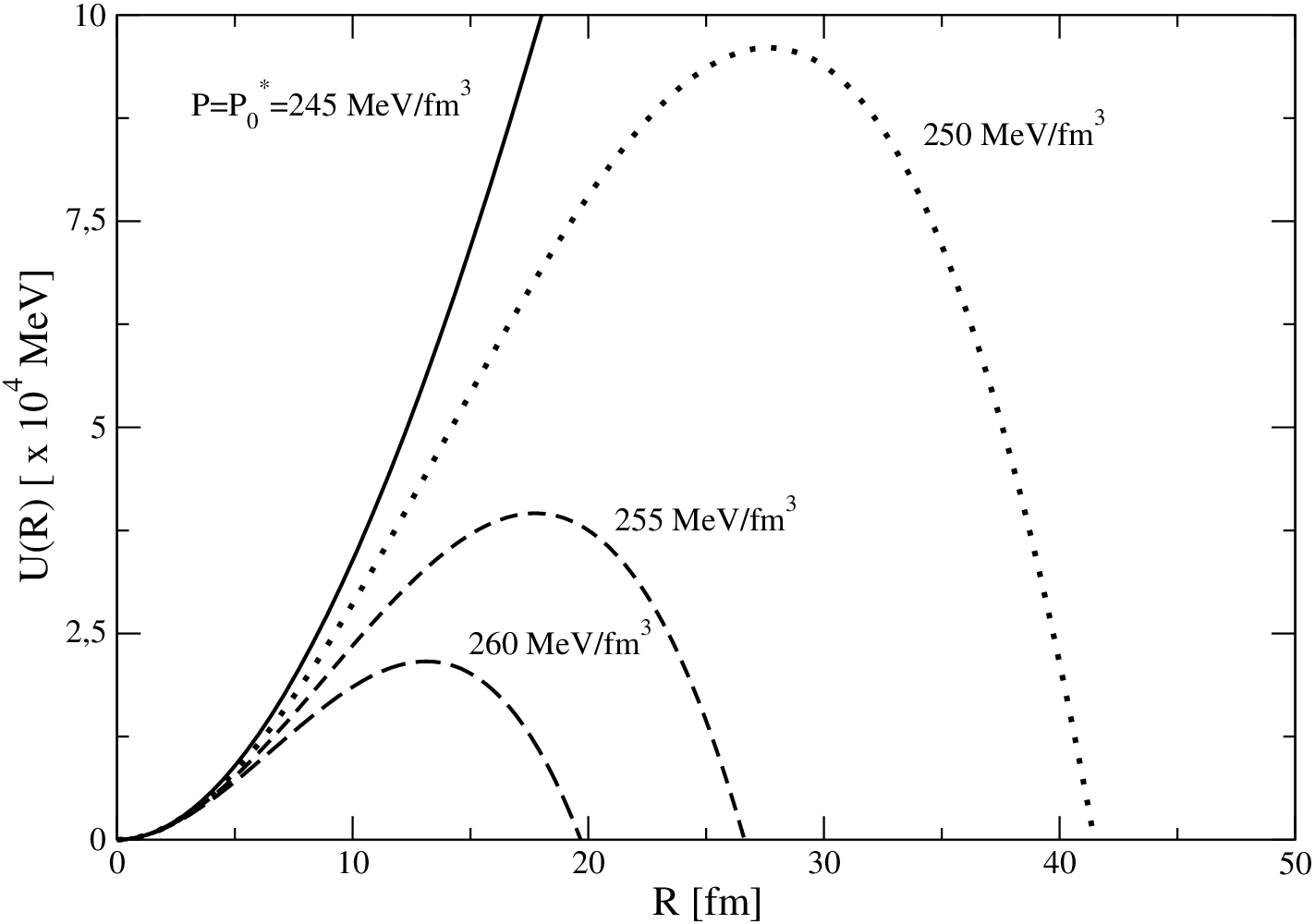}  
\caption{Potential energy of the QM drop as a function of the radius of the drop for 
several pressures above $P_0^*$.  The hadronic and Q$^*$ phases are described with 
the same EoS employed in Fig. 1. 
The surface tension $\sigma$ is taken equal to $30$ MeV/fm$^2$ 
and the curvature coefficient $\gamma$ is taken equal to zero.  }
\label{fig:fig3}
\end{figure}

As an illustrative example, we plot in Fig.\ \ref{fig:fig3} the potential energy $U(\R)$ 
for the formation of a quark matter droplet for different values of the stellar 
central pressure $P_c$ above the static transition point $P_0^*$.  
The curves in Fig.\ \ref{fig:fig3}  are relative to  a given set of EOS 
for the two phases of dense matter (see figure caption), to a fixed value of  
the surface tension $\sigma$ and to the case $\gamma = 0$.  
  
As expected the potential barrier is lowered as central pressure increases.   

\begin{figure}
%%\resizebox{\hsize}{!}{\includegraphics{figure4.eps}}
%%\plotone{figure4.eps}
\plotone{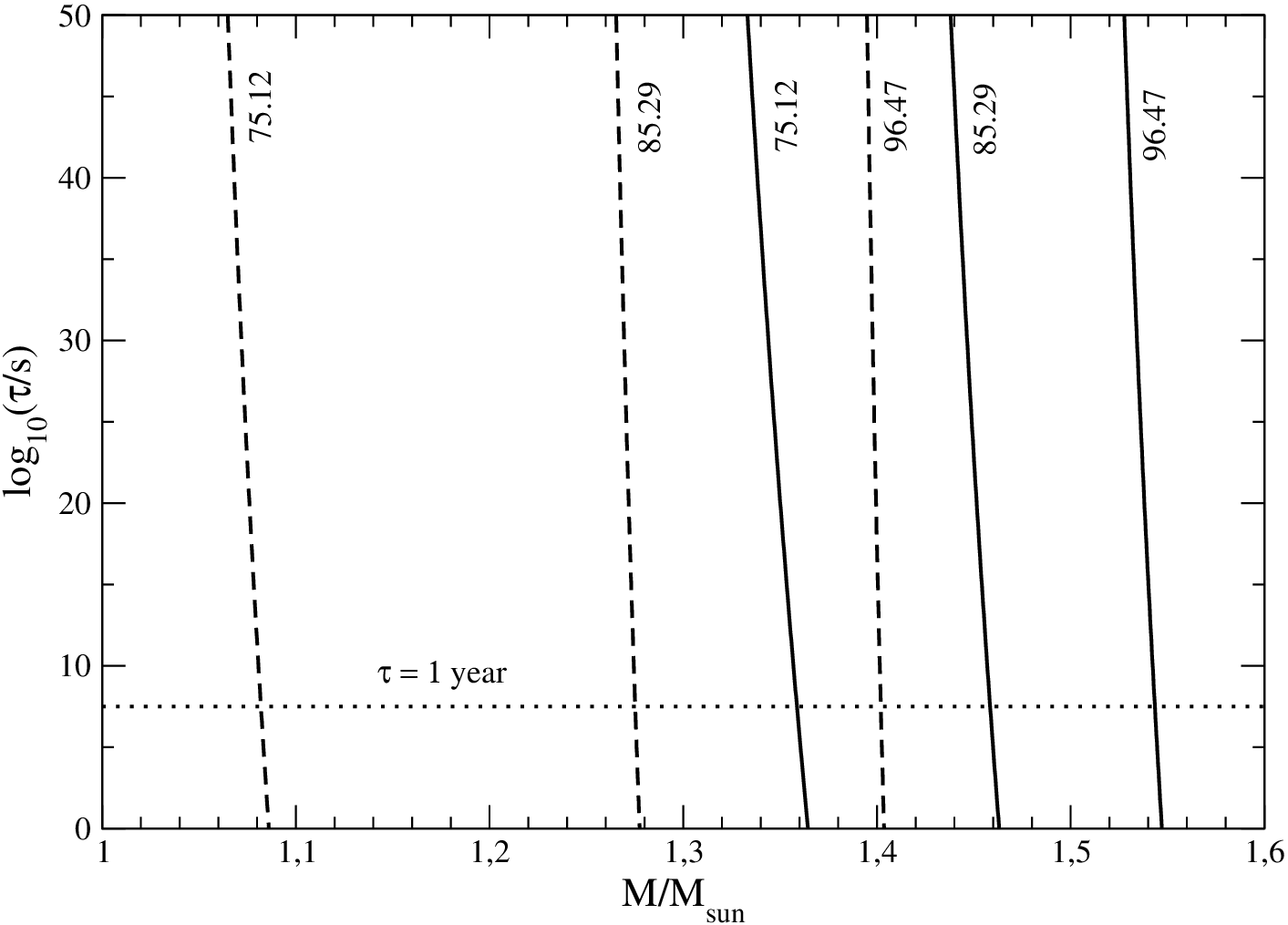}  
\caption{Nucleation time as a function of the maximum gravitational mass of the 
hadronic star. Solid lines correspond to a value of $\sigma=30$ MeV/fm$^2$ whereas 
dashed ones are for $\sigma=10$ MeV/fm$^2$. In both cases we take $\gamma = 0$.   
The nucleation time correspondig to one year  is shown by the dotted horizontal line. 
The different values of the bag constant (in units of MeV/fm$^3$) are plotted next 
to each curve. The hadronic phase is described with the GM1 model.}
\label{fig:fig4} 
\end{figure}

\begin{figure}
%\resizebox{\hsize}{!}{\includegraphics{figure5.eps}}
%%\plotone{figure5.eps} 
\plotone{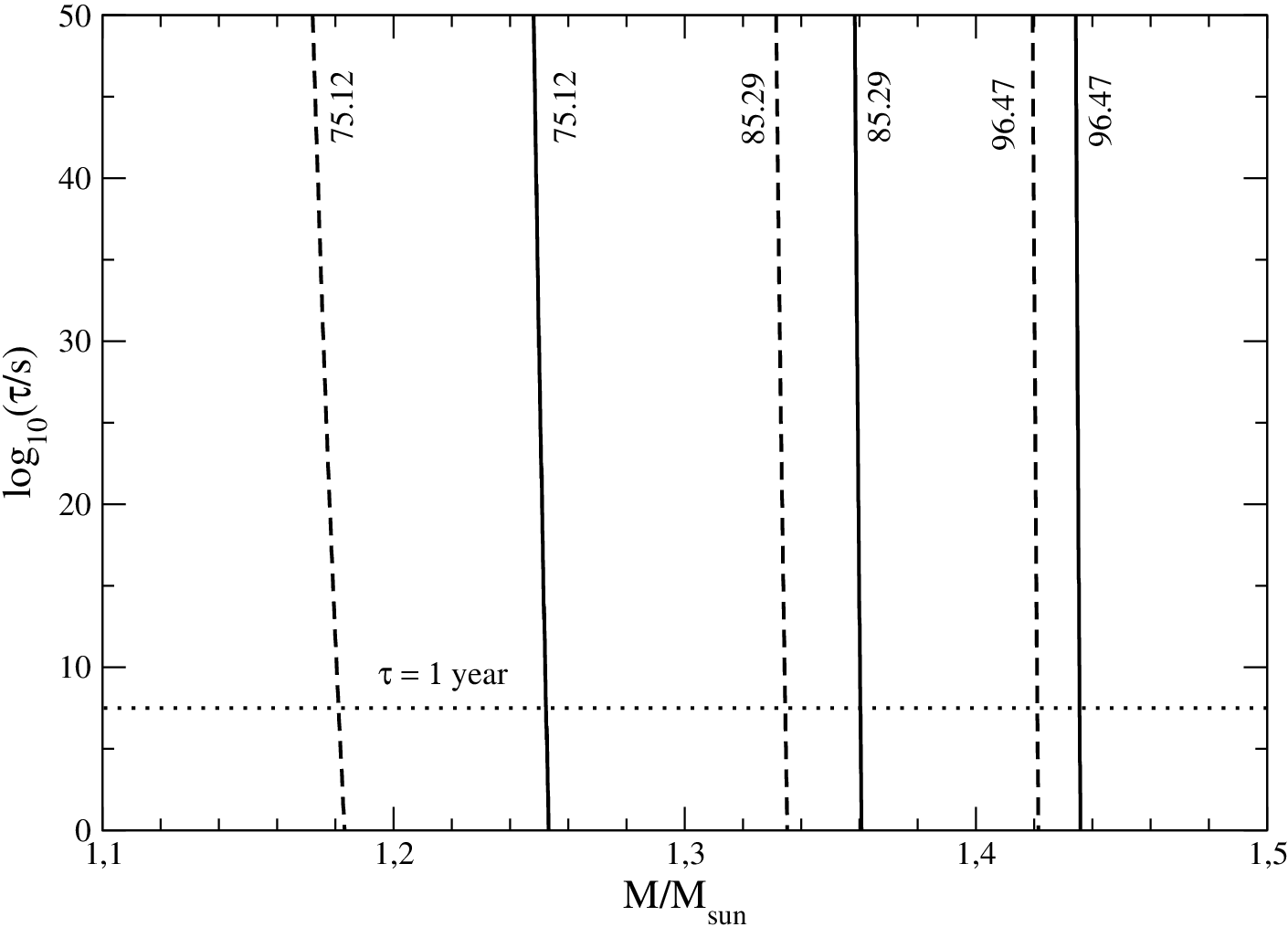}   
\caption{Same as Fig. 4 for the GM3+Bag model.}
\label{fig:fig5}
\end{figure}

The nucleation time $\tau$, {\it i.e.,} the time needed to form a critical droplet 
of deconfined quark matter,  can be calculated for different values of the stellar 
central pressure $P_c$ which enters in the expression of the energy barrier 
in Eq.\ (\ref{eq:eq7}). The nucleation time can be plotted as a function of the 
gravitational mass $M_{HS}$ of the HS corresponding to the given value of the 
central pressure, as implied by the solution of the Tolmann-Oppeneimer-Volkov 
equations for the pure Hadronic Star sequences. 
The results of our calculations are reported in  Fig.\ \ref{fig:fig4} and 
in Fig.\ \ref{fig:fig5} which are relative respectively to the GM1 and GM3 EOS 
for the hadronic phase.   Each curve refers to a different value of the bag constant 
and the surface tension.   
As we can see,  from the results in Fig.s\  \ref{fig:fig4} and  \ref{fig:fig5}, 
a metastable  hadronic star can have a mean-life time many orders of magnitude 
larger than the age of the universe 
$T_{univ} = (13.7 \pm 0.2) \times 10^{9}$~yr $ = (4.32 \pm 0.06) \times 10^{17}$~s  
(Spergel et al. 2003).    As the star accretes a small amount of mass  
(of the order of a few per cent of the mass of the sun),  
the consequential increase of the central pressure lead to a huge 
reduction of the nucleation time and, as a result, to a dramatic reduction 
of the HS {\it mean-life time}.  

To summarize, in the present  scenario pure hadronic stars having  a central 
pressure larger than the static transition pressure  for the formation 
of the Q*-phase are metastable to the ``decay'' (conversion) to a more compact 
stellar configuration in which deconfined quark matter is present 
({\it i.e.,} HyS or SS). These metastable HS  have a {\it mean-life time}  which is 
related to the nucleation time to form the first critical-size drop of deconfined 
matter in their interior (the actual  {\it mean-life time} of the HS will depend on the 
mass accretion or on the spin-down rate which modifies the nucleation time via  
an explicit time dependence of the stellar central pressure).  
We define as  {\it critical  mass} $M_{cr}$ of the metastable HS,  
the value of the  gravitational mass for which the nucleation time is equal to one year: 
$M_{cr} \equiv M_{HS}(\tau = 1 {\rm yr})$.   
Pure hadronic stars with $M_H > M_{cr}$ are very unlikely to be observed.  
$M_{cr}$  plays the role of an {\it effective maximum mass} 
for the hadronic branch of compact stars (see the discussion in subsection 3.1).   
While the Oppenheimer--Volkov maximum mass $M_{HS,max}$ 
(Oppenheimer \& Volkov 1939)  is determined 
by the overall stiffness of the EOS for hadronic matter, 
the value of $M_{cr}$  will depend in addition on the bulk properties of the EOS 
for quark matter and on the properties at the interface between 
the confined and deconfined phases of matter ({\it e.g.,} the surface tension $\sigma$).

To explore how the outcome of our scenario depends on the details of the stellar 
matter EOS,  we have considered two different parameterizations (GM1 and GM3) 
for the EOS of the hadronic phase, and we have varied the value of the bag constant $B$. 
Moreover, we have considered two different values for the surface tension: 
$\sigma = 10$~MeV/fm$^2$ and $\sigma = 30$~MeV/fm$^2$ 
(taking $\gamma = 0$ in both cases).   
These results are summarized in Tab. 1 (GM1 EOS) and in Tab. 2 (GM3 EOS)

In Fig.\ \ref{fig:fig6} and \ref{fig:fig7},  we show the MR curve for pure HS within 
the GM1 and GM3 models for the EOS of the hadronic phase, and that for hybrid stars 
or strange stars for different values of the bag constant $B$.     
The configuration marked with an asterisk on the hadronic MR curves represents 
the hadronic star for which the central pressure is equal to $P_0^*$.   
The full circle on the hadronic star sequence represents the critical mass configuration, 
in the case  $\sigma = 30$ MeV/fm$^2$ and $\gamma = 0$.     
The  full circle on the HyS (SS) mass-radius curve represents the hybrid (strange) star  
which is formed from the conversion of the hadronic star with $M_{HS} = M_{cr}$. 
We assume (Bombaci \& Datta 2000)  that during the stellar conversion process 
the total number of baryons in the star (or in other words the stellar baryonic mass)  
is conserved.  Thus the total energy liberated in the stellar conversion is given by 
the difference between the gravitational mass of the initial hadronic star    
($M_{in} \equiv M_{cr}$)  
and that of the final hybrid or strange stellar configuration with 
the same baryonic mass ($M_{fin} \equiv M_{QS}(M^b_{cr}) \,$):  
\begin{equation}
          E_{conv} = (M_{in} - M_{fin}) c^2  \, . 
\label{eq:eq11}
\end{equation}

The stellar conversion process, described so far, will start to populate 
the new branch of quark stars (the part of the QS sequence plotted as a continuous 
curve in  Fig.s\ \ref{fig:fig6} and  \ref{fig:fig7}). 
Long term accretion on the QS can next produce stars with 
masses up to the limiting mass $M_{QS,max}$ for the quark star configurations.

As we can see from the results reported in Tab. 1 and 2,  within the present 
model for the EOS, we can distinguish several ranges for the value 
of the bag constant, which gives a different astrophysical output for our scenario. 
To be more specific, in Fig.\ \ref{fig:fig8}  we plot the maximum mass of the 
QS sequence, the critical mass and the corresponding final mass $M_{fin}$ 
as a function of $B$, in the particular case of the GM3 model for the EOS of the 
hadronic phase and taking $\sigma = 30 {\rm MeV/fm}^2$  and $\gamma = 0$.     
Let us start  the following discussion from ``high'' values of $B$  down to the 
minimum possible value $B^{V}$ ($\sim 57.5 {\rm MeV/fm}^3$ for $\alpha_s = 0$)  
for which atomic nuclei will be unstable to the decay to a drop 
of deconfined {\it u,d} quark matter (non-strange QM) (Farhi \& Jaffe 1984).  

{\bf (1)~}  $B > B^I \,$.~~
These ``high'' values of the bag constant do not allow the quark deconfinement 
to occur in the maximum mass hadronic star either.    
Here  $B^I$ denotes the value of the bag constant for which 
the central density of the maximum mass hadronic star is equal to 
the critical density for the beginning of the mixed quark-hadron phase.     
For these values of $B$,  all compact stars are pure hadronic stars.    

{\bf (2)~}  $B^{II}  < B <  B^{I} \,$.~~ Now, in addition to pure HS, there is a new branch 
of compact stars, the hybrid stars;  but the nucleation time $\tau(M_{HS,max})$ 
to form a droplet of Q*-matter in the maximum mass hadronic star,  
is of the same order or much larger than the age of the Universe.     
Therefore, it is extremely unlikely to populate the hybrid star branch. 
Once again, the compact star we can observe are, in this case, pure HS.  

{\bf (3)~}  $B^{III}  < B <  B^{II} \,$.~~ 
In this case, the critical mass for the pure hadronic star sequence is less than 
the maximum mass for the same stellar sequence, {\it i.e.,}  $M_{cr}  < M_{HS,max}$.   
Nevertheless (for the present EOS model),  the baryonic mass $M^{b}(M_{cr})$ of 
the hadronic star with the critical mass is larger than the maximum baryonic mass 
$M^b_{QS,max}$ of the hybrid star sequence.  
In this case,  the formation of a critical size droplet of deconfined matter in the core of 
the hadronic star with the critical mass, will trigger off a stellar conversion process 
which will produce, at the end, a black hole  
(see cases marked as ``BH'' in Tab. 1 and Tab. 2).      
As in the previous case,  it is extremely unlikely to populate the hybrid star branch. 
The compact star predicted by these EOS models  are pure HS.  
Hadronic stars with a gravitational  mass in the range 
$M_{HS}(M^b_{QS,max})  < M_{HS}  < M_{cr}$  
(where $M^b_{QS,max}$  is the baryonic mass of 
the maximum mass configuration for the hybrid star sequence)  
are metastable with respect to a conversion to a black hole.     
 
{\bf (4)~}  $B^{IV}  < B <  B^{III} \,$.~~ 
For these values of $B$ one has   $M_{cr}  <  M_{HS}(M^b_{QS,max})$ . 
There are now two different branches of compact stars: 
pure hadronic stars with $M_{HS} < M_{cr}$,   and 
hybrid stars with $M_{QS}(M^b_{cr})  < M_{QS}  < M_{QS,max} \,$    
(here $M_{QS}(M^b_{cr}) \equiv M_{fin}$ is the gravitational mass of the hybrid star 
with the same baryoinic mass  of the critical mass hadronic star).     

{\bf (5)~}  $B^{V}  < B <  B^{IV} \,$.~~ 
Finally, as  $B$ falls below the value $B^{IV}$,  the Bodmer-Witten 
hypothesis starts to be fulfilled.  Now the stable quark stars formed 
in the stellar conversion process are strange stars.

%%%  REV.VERSION 01/04 %%%%%%%%
To examine the influence of the curvature energy on the quantum nucleation 
process, we have performed some  calculations of the nucleation time taking 
$\gamma \neq 0$ in Eq.\ (\ref{eq:eq7}).  We have used the two values, 
$\gamma = 10$~MeV/fm and   $\gamma = 20$~MeV/fm,  according to the 
existing estimates of the curvature coefficient in the case of the quark-hadron 
phase transition (Madsen 1993; Horvath 1994; Olesen \& Madsen 1994).   
As expected, for a fixed value of the stellar central pressure, the curvature term 
suppresses quark matter nucleation.  
As a consequence, the  critical mass $M_{cr}$  and the total energy released 
in the stellar conversion process  are increasesd with respect to  the case 
$\gamma = 0$, as it can be seen looking at the results reported in Tab. 3.     
%%%%%%%%%%%%%%%%%%%%%%%%%    

\subsection{The limiting mass of compact stars} 

The possibility to have metastable hadronic stars, together with the feasible 
existence of two distinct families of compact stars, demands an extension of the 
concept of maximum mass of a ``neutron star'' with respect to 
the {\it classical} one introduced by Oppenheimer \& Volkoff (1939).    
Since metastable HS with a ``short'' {\it mean-life time} are very unlikely to be observed,  
the extended concept of maximum mass must be introduced in view of the comparison 
with  the values of the mass of compact stars deduced from direct astrophysical 
observation.  
Having in mind this operational definition,  we call  {\it limiting mass} of a compact star, 
and denote it as $M_{lim}$, the physical quantity defined in the following way: 

\noindent 
({\it a})  if the nucleation time $\tau(M_{HS,max})$  associated to the maximum mass 
configuration for the hadronic star sequence is of the same order or much larger  than  
the age of the universe $T_{univ}$,   then 
\be
         M_{lim}  =  M_{HS,max} \, ,  
\ee
in other words, the limiting mass in this case coincides with the Oppenheimer--Volkoff 
maximum mass for the hadronic star sequence. 

\noindent 
({\it b}) If the critical mass $M_{cr}$  is smaller than $M_{HS,max}$  
({\it i.e.}  $\tau(M_{HS,max}) < 1$~yr), 
thus the limiting mass for compact stars is equal to the largest value between the 
critical mass for the HS and the maximum mass for the quark star (HyS or SS) sequence 
\be
         M_{lim} =   max \big[M_{cr} \, ,  M_{QS,max} \big] \, .
\ee

\noindent 
({\it c}) Finally, one must consider an ``intermediate'' situation for which      
                          $1 {\rm yr} <  \tau(M_{HS,max})  <  T_{univ}$.  
As the reader can easely  realize, now 
\be
         M_{lim} =   max \big[M_{HS,max} \, ,  M_{QS,max} \big] \, , 
\ee
depending on the details of the EOS which could give $M_{HS,max} > M_{QS,max}$ 
or vice versa.  

In   Fig.\ \ref{fig:fig9}, we show the limiting mass  $M_{lim}$ calculated in the case of the  
GM1+Bag model (dashed line) and in the case of the GM3+Bag model (continuous line)  
as a function of the bag constant $B$.  
In the same figure, we compare our theoretical determination for 
$M_{lim}$ with some of the ``measured'' masses of compact stars in radio pulsar 
binaries (Thorsett \& Chakrabarty 1999) and for the compact stars 
Vela X-1  (Quaintrell et al.  2003) and   Cygnus X-2  (Orosz \& Kuulkers 1999).

\section{Mass-to-radius  ratio and  internal constitution of compact stars}  

An accurate measure of the radius and the mass of an individual 
``neutron star'' will represent the key to open the {\it safety deposit}  
which contains the secrets of the internal constitution of these puzzling 
astrophysical bodies and to discriminate between different models for 
the equation of state of dense hadronic matter.  
Unfortunately such a crucial information is still not available. 
A decisive step in such a direction has been done thanks to the instruments on board 
of the  last generation of X-ray satellites. These  are providing a large amount 
of fresh and accurate  observational data, which are giving us  the possibility to 
extract very tight constraints on the radius and the mass for some compact stars.  
  
 The analysis of different astrophysical  phenomena associated with 
compact X-ray sources,  seems to indicate in some case the existence of 
neutron stars with ``large'' radii in the range of 12 -- 20 km 
and in some other cases the  existence of  compact stars with ``small'' radii in 
the range of 6 -- 9 km (Bombaci 1997;  Li et al. 1999a; 
Poutanen \& Gierlinski 2003; Bombaci 2003).    
Clearly, this possibility is a natural outcome of our scenario, where two different 
families of compact stars, the pure hadronic stars and the quark stars (HyS or SS), 
may exist in the universe.  

In the following of this section, we will consider some of the most  recent  
constraints on the mass-to-radius ratio for compact stars extracted from the 
observational data for a few X-ray sources, 
and we will try make an interpretation of these results within our scenario.    

In Fig.\ \ref{fig:fig10}, we report the radius an the mass of the compact star 
RX J1856.5-3754 inferred by Walter \& Lattimer (2002) 
(see also Kaplan et al. 2002)  from the fit of the full spectral energy distribution 
for this isolated radio-quite ``neutron star'', after a revised parallax determination 
(Kaplan et al. 2002) which  implies a distance to the source of $117 \pm 12$~pc.  
Comparing the mass-radius box for  \rx18  reported in Fig.\ \ref{fig:fig10} 
with the theoretical determination of the MR relation for different 
equations of state, one concludes that \rx18 could be  (see {\it e.g.} Fig. 2 
in Walter \& Lattimer, 2002) either an hadronic star or an hybrid or 
strange star (see also Drake et al. 2002).    

Next we consider the compact star in the low mass X-ray binary 4U~1728-34.   
In a very recent paper Shaposhnikov  et al. (2003)  (hereafter STH)  
have analyzed a set of 26 Type-I X-ray bursts  for  this source. 
The data were collected by the Proportional Counter Array 
on board of the Rossi  X-ray Timing Explorer (RXTE) satellite. 
For the interpretation of these observational  data 
Shaposhnikov  et al.  2003 used a model of the X-ray burst 
spectral formation developed by Titarchuk (1994)  and  
Shaposhnikov \& Titarchuk  2002.    
Within this model, STH were able to extract very stringent constrain on 
the radius and the mass of the compact star in this bursting source.  
The radius and mass for 4U~1728-34, extracted by STH for different best-fits 
of the burst data, are depicted in  Fig.\ \ref{fig:fig10} 
by the filled squares.   Each of the four MR  points is relative to a different value of the 
distance to the source (d = 4.0, 4.25, 4.50, 4.75 kpc, for the fit which produces 
the smallest values of the mass, up to the one which gives the largest mass).   
The error bars on each point represent the error contour for 90\% confidence level.  
It has been pointed out (Bombaci 2003) that the semi-empirical MR relation for 
the compact star in 4U~1728-34 obtained by STH  is not compatible with models 
pure hadronic stars,  while it is consistent with strange stars or hybrid stars.   

Assuming \rx18 to be a pure hadronic star and 4U~1728-34 an hybrid or a strange 
star, we see from our results plotted in   Fig.\ \ref{fig:fig10}, 
that this possibility can be realized as a natural consequence of our scenario. 
Thus, we find that the existence of quark stars (with ``small'' radii) does not exclude 
the possible existence of pure hadronic stars (with ``large'' radii), and {\it vice versa}.   

\begin{figure}
%%\resizebox{\hsize}{!}{\includegraphics{figure6.eps}}
%%%\plotone{figure6.eps}
\plotone{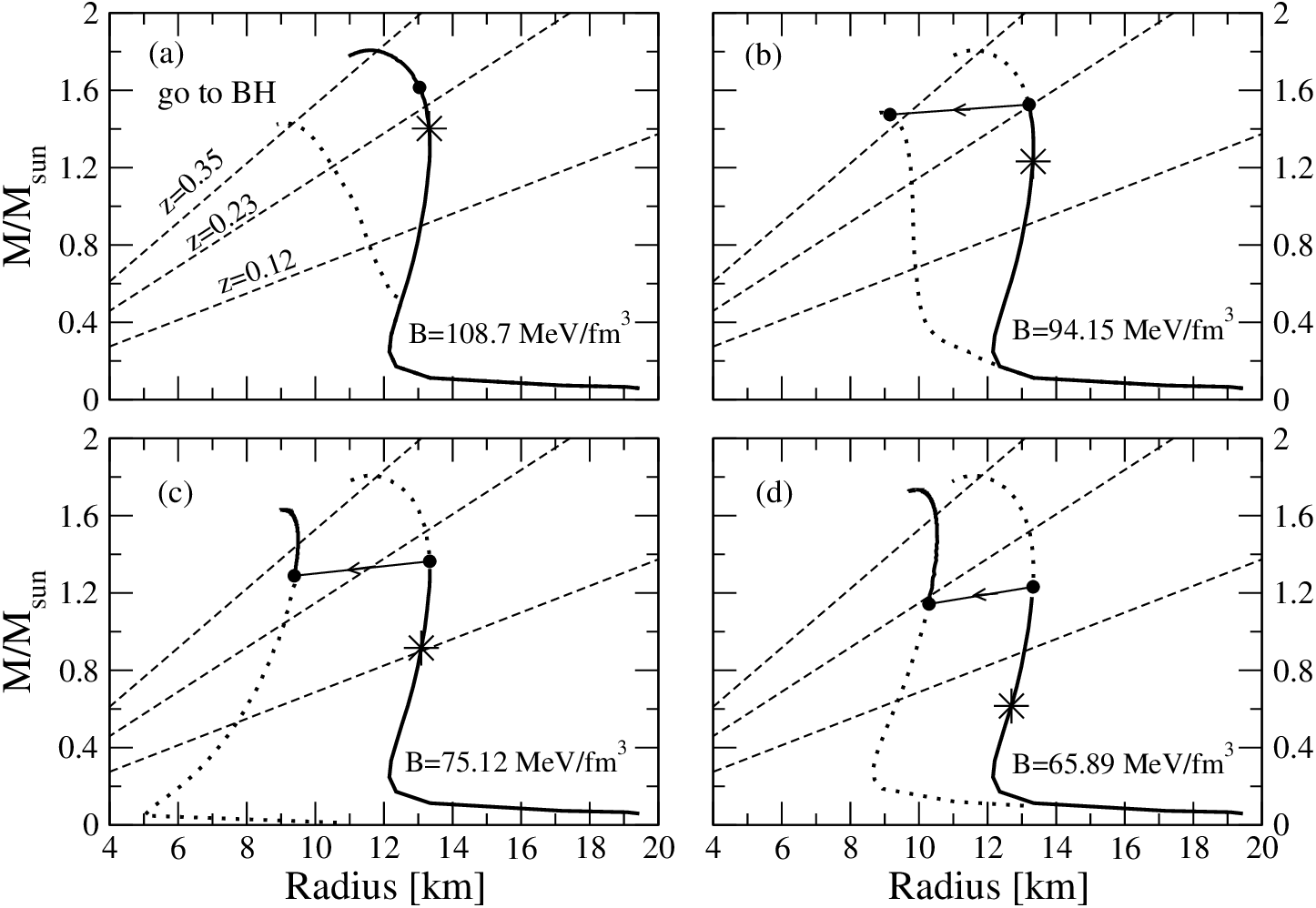}   
\caption{Mass-radius relation for a pure HS described within the GM1 model and 
that of the HyS or SS configurations for several values of the Bag constant and 
$m_s=150$ MeV and $\alpha_s=0$. 
The configuration marked with an asterisk represents 
in all cases the HS for which the central pressure is equal to $P_0^*$. 
The conversion process of the HS, with a gravitational mass equal to $M_{cr}$, into
a final HyS or SS is denoted by the full circles connected by an arrow. 
In all the panels $\sigma$ is taken equal to $30$ MeV/fm$^2$ 
and $\gamma = 0$.     
The dashed lines show the gravitational red shift deduced for the X-ray compact 
sources \exo07 ($z = 0.35$) and \1e12 ($z =$ 0.12 -- 0.23). }
\label{fig:fig6}
\end{figure}

\begin{figure}
%%\resizebox{\hsize}{!}{\includegraphics{figure7.eps}}
%%\plotone{figure7.eps}
\plotone{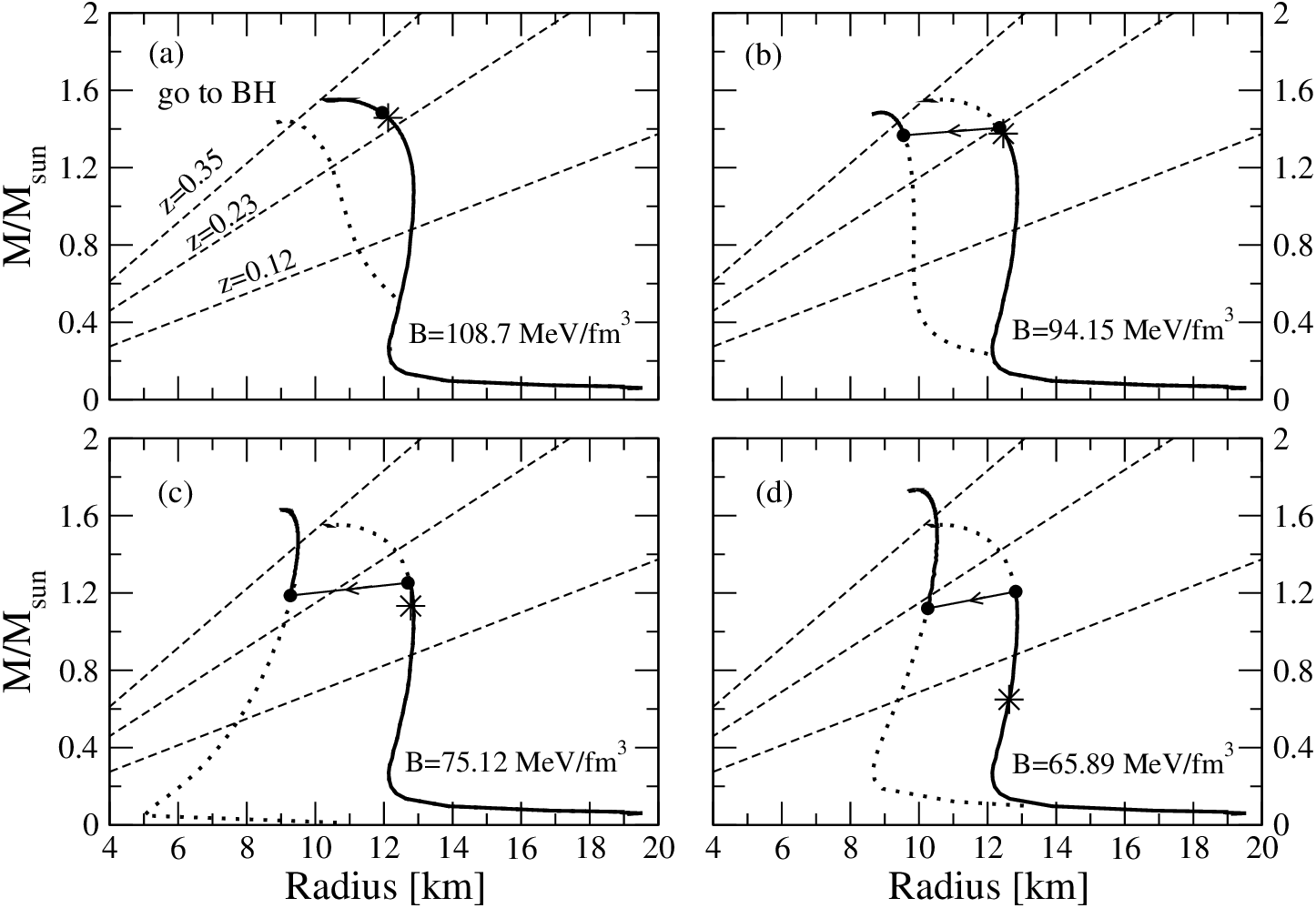}   
\caption{Same as Fig. 6 for the GM3+Bag model.}
\label{fig:fig7}
\end{figure}

\begin{figure}
%%\resizebox{\hsize}{!}{\includegraphics{figure8.eps}}
%%%\plotone{figure8.eps}
\plotone{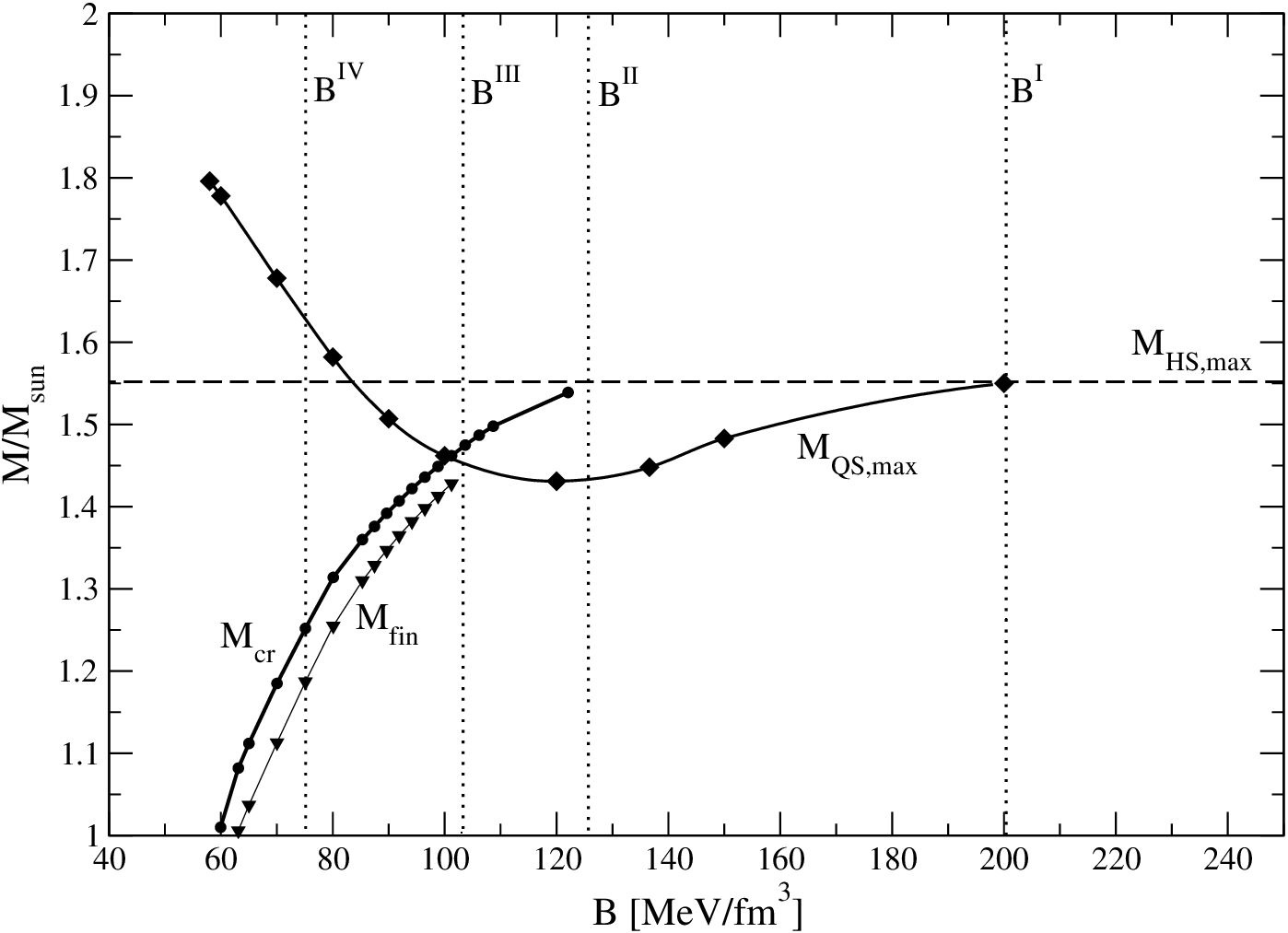}  
\caption{The maximum mass $M_{QS,max}$ for the quark star configuartions (HS or SS),  
the critical mass $M_{cr}$ and the mass $M_{fin}$ of the stable QS to which it evolves 
are plotted as a function of the bag constant $B$.  
The vertical doted lines labelled $B^I$ -- $B^{IV}$  mark the boundary of different 
ranges of the bag constant which give a different astrophysical output for our scenario, 
as discussed in the text. 
The dashed horizontal line gives the value oh the maximum mass for the pure 
hadronic star sequence.  
All the results are relative to the GM3 model for the EOS for the hadronic phase, 
the surface tension $\sigma$ is taken equal to $30$ MeV/fm$^2$.  }

\label{fig:fig8}
\end{figure}

\begin{figure}
%%\resizebox{\hsize}{!}{\includegraphics{figure9.eps}}
%%\plotone{figure9.eps}
\plotone{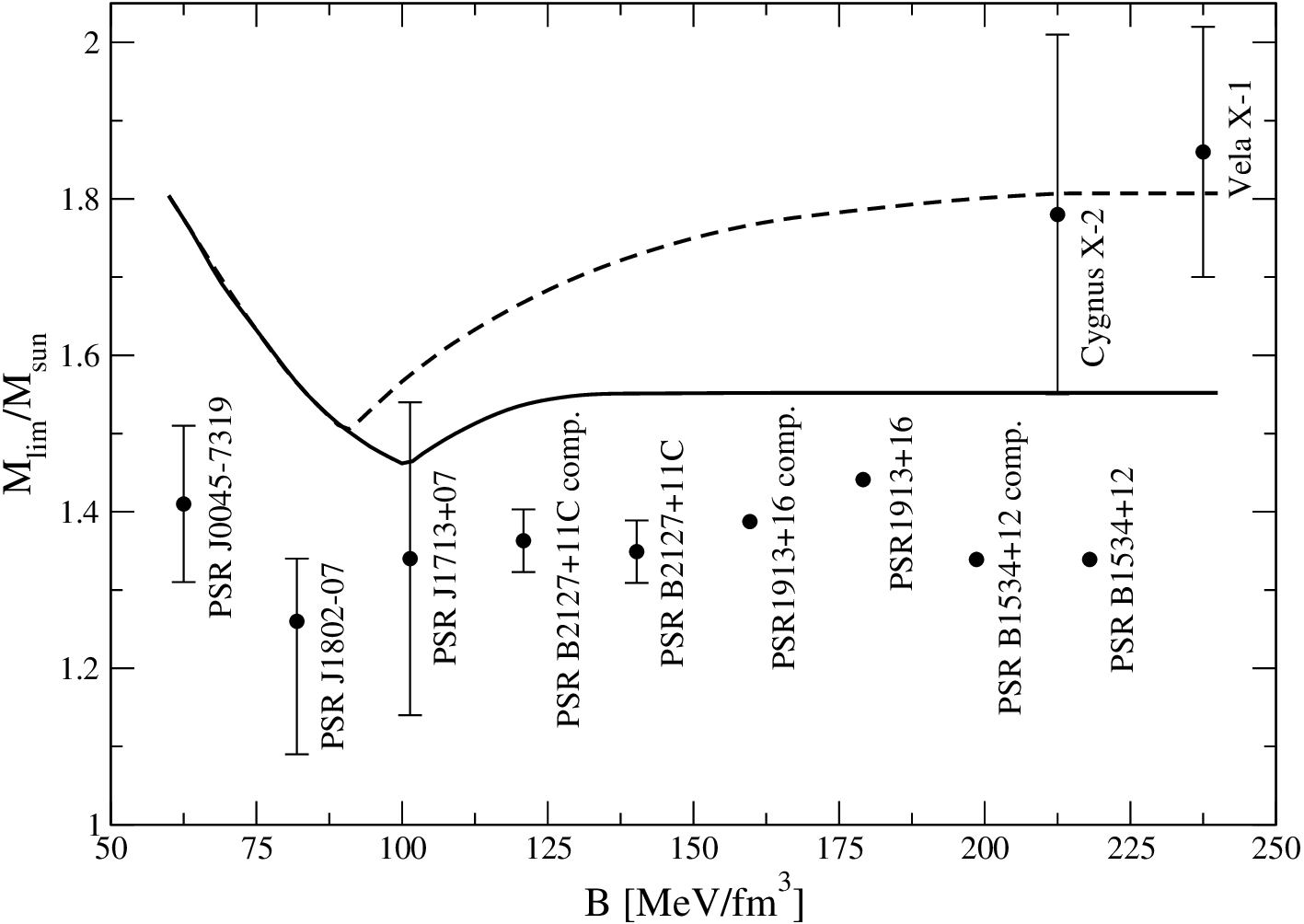}  
\caption{The limiting (gravitational) mass $M_{lim}$,  according to generalized 
definition given in the present work, is plotted as a function of the Bag constant.   
Solid (dashed) lines show the results for the GM3+Bag (GM1+Bag) model. 
In both cases we take  $\sigma  = 30$ MeV/fm$^2$  and $\gamma = 0$.  
The values of some ``measured'' masses of compact stars in radio pulsars and 
in Vela X-1 and Cygnus X-2 are also reported for comparison.}
\label{fig:fig9}
\end{figure}

\begin{figure}
%\resizebox{\hsize}{!}{\includegraphics{figure10.eps}}
%%%\plotone{figure10.eps}
\plotone{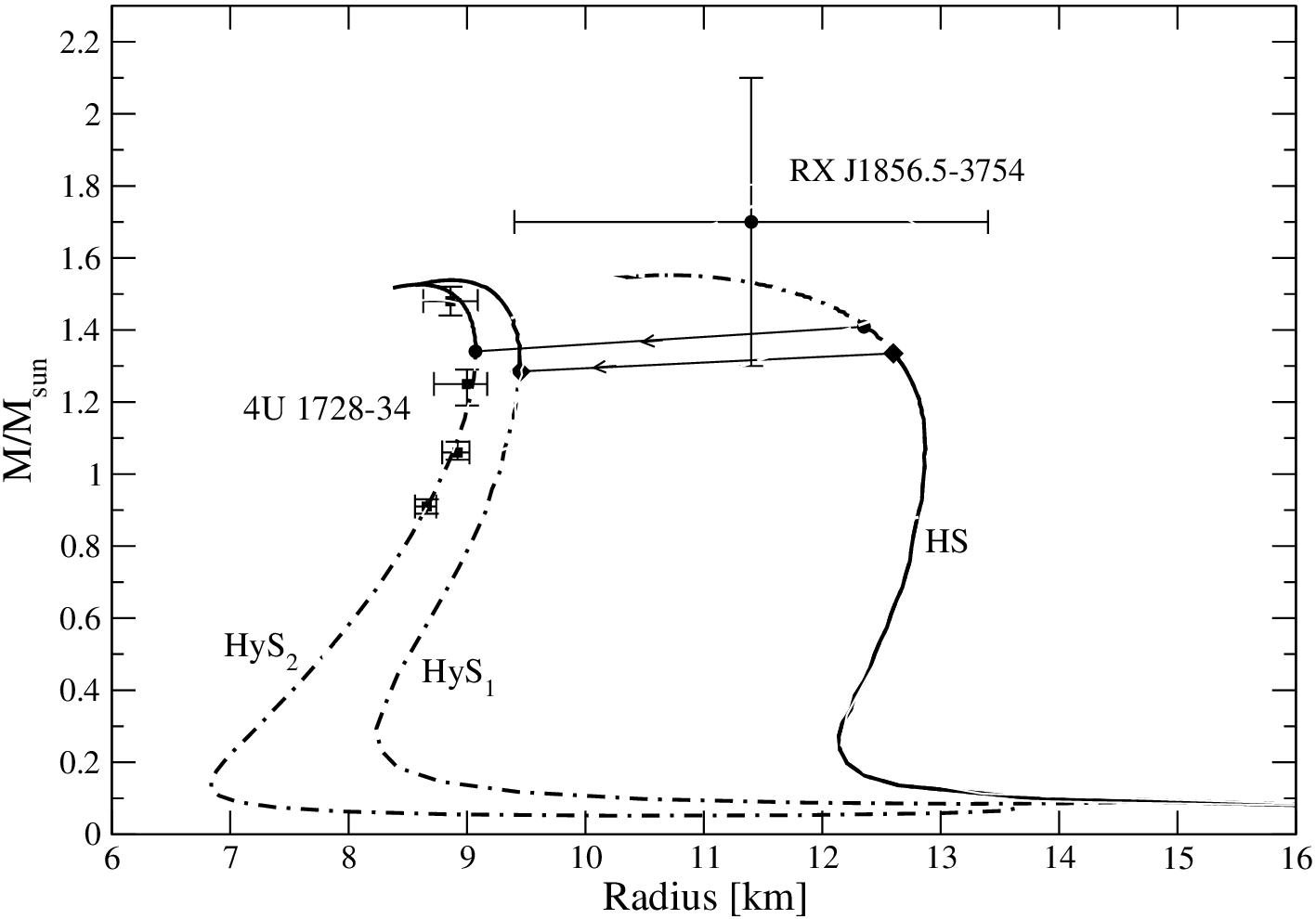}  
\caption{The radius and the mass for  RX J1856.5-3754 (full circle with error 
bars labelled  RX J1856.5-3754) obtained  by Walter \& Lattimer (2002) 
from fitting  the multi-wave lenght spectral energy distribution. 
The radius and mass for 4U~1728-34, extracted by 
Shaposhnikov {\it et al.}  (\cite{shapo})  for different best-fits 
of the X-ray burst data, is shown by the filled circles with error bars  
(error contour for 90\% confidence level).     
The  curves labeled HS  reprersents the MR relation for pure hadronic star 
with the GM3 equation of state. 
The curves labeled HyS$_1$ and HyS$_2$ are the MR curves for hybrid stars 
with the GM3+Bag model EOS, for 
$B= 85.29 {\rm MeV/fm}^3$ and $m_s = 150 $ MeV (HyS$_1$), and 
$B= 100 {\rm MeV/fm}^3$ and $m_s = 0 $ MeV (HyS$_2$).  
The full circles and diamonds on the MR curves represent the 
critical mass configuration (symbols on the HS curve) and the corresponding 
hybrid star configurations after the stellar conversion process (symbols on 
the HyS$_1$ and HyS$_2$ curves). 
}  
\label{fig:fig10}
\end{figure}

Decisive informations on the mass-to-radius ratio can be provided  
by measuring the  gravitational redshift of lines in the spectrum emitted 
from the compact star atmosphere.  
Very recently, redshifted spectral lines features have been reported for two different 
X-ray sources (Cottam et al. 2002; Sanwal et al. 2002).   
The first of these sources is the compact star  in the low mass X-ray binary 
\exo07.  Studing the spectra of 28 type-I X-ray bursts  in \exo07,  
Cottam et al. (2002) have found absorption spectral line features, 
which they identify as signatures of Fe~XXVI (25-time ionized hydrogen-like Fe) 
and  Fe~XXV from the $n = 2 \rightarrow 3$ atomic transition, 
and of O~VIII ($n = 1 \rightarrow 2$ transition).  
All of these lines are redshifted, with a unique value of the redshift $z= 0.35$. 
Interpreting the measured redshift as due to the strong gravitational field   
at the surface of the compact star (thus neglecting general relativistic effects due 
to stellar rotation on the spectral lines (\"Oezel \& Psaltis 2003) ),   
one obtains a relation for the stellar mass-to-radius ratio:   
\be
              M/M_\odot   = \Big( 1 - {{1}\over{(z+1)^2}} \Big) R/R_{g\odot} \, , 
\ee
($R_{g\odot} = 2 GM_\odot/c^2 = 2.953$~km)   
which is reported in  Fig.s\ \ref{fig:fig6} and  \ref{fig:fig7}  as a dashed line 
labeled $z= 0.35$.   
Comparing with the theoretical MR relations for differente EOS 
(see {e.g.}  Fig.\ \ref{fig:fig6},, and also Xu 2003)  it is clear that  all  three possible 
families of compact stars discussed in the present paper are completely 
consistent with a redshitf  $z=0.35$.  

The second source for which it has been claimed the detection of redshifted 
spectral lines is 1E~1207.4-5209,   a radio-quite compact star      
located in the center of the supernova remnant PSK~1209-51/52.   
1E~1207.4-5209  has been observed by the Chandra X-ray observatory. 
Two absorption features have been detected in the source spectrum and 
have been interpreted  (Sanwal et al. 2002)  as spectral lines associated 
with atomic transitions of once-ionized helium in the atmosphere of a 
strong magnetized ($ B\sim 1.5 \times 10^{14}$~G) compact star.  
This interpretation gives for the gravitational redshift at the star surface 
$z = $ 0.12 -- 0.23  (Sanwal et al. 2002),   
which is reported  in  Fig.s\ \ref{fig:fig6} and  \ref{fig:fig7}  by the two dashed lines 
labeled $z= 0.12$ and $z=0.23$.     

A different interpretation of similar data, collected by the XMM-Newton 
satellite, has been recently given by  Bignami et al.  (2003),   
who interpreted the absorption features in the spectrum of \1e12 as 
electron cyclotron lines in the stellar magnetic field.  
Within this interpretation and assuming the gravitational redshift of a 
``canonical neutron star'' with $M=1.4 M_\odot$ and $R =10$~km 
Bignami  et al. (2003)  derived a magnetic field strength 
$B = 8 \times 10^{10}$~G for the compact star in \1e12. 

The two values of the stellar magnetic field inferred in the above quoted 
papers  (Sanwal  et al. 2002;   Bignami et al. 2003)   
are in disagreement to each other and in disagreement with the 
$B$ field deduced from the \1e12 timing parameters ($P$ and $\dot{P}$), 
which give $B = $ (2 -- 3) $\times 10^{12}$~G within  the rotating magnetic 
dipole  model.     
However, the latter value of the magnetic field strenght presents serious 
problems  since the current values of the timing parameters implies 
a characteristic pulsar age $\tau_c = P/(2\dot{P}) \sim 4.8 \times 10^5$~yr 
(Bignami et al. 2003) 
which is not compatible with the age $\tau_{SNR} =  (3$--$20) \times 10^3$~yr 
(Roger et al. 1988).         
Clearly this source needs a more accurate study before any final and 
unambiguous interpretations of the observed spectral features can be drawn.   
Here, we will assume that the interpretation of the spectral feature given by  
Sanwal et al.  (2002) and  by Cottam et al. (2002) is  correct.  
In that case, how it is possible to reconcile the gravitational redshift 
$z = $012--0.23 for 1E~1207.4-5209 with that ($z = 0.35$) deduced for EXO 0748-676?  
Within the commonly accepted view, in which there exist in nature only one family 
of compact stars (the ``neutron stars''),  different values of the gravitational 
redshift could be  a consequence of a different mass of  the two stars. 
In our scenario, we can give a different interepretation:  
1E~1207.4-5209 is a pure hadronic star whereas EXO 0748-676 is an hybrid star 
or a strange star. This is illustrated in Fig.s\ \ref{fig:fig6} and \ref{fig:fig7}  
by comparing our calculated MR relations with the redshifts deduced for 
the two compact X-ray sources.      

\section{Quark Deconfinement Nova and GRBs}  

A large variety of  observational data are  giving a mounting evidence that 
``long-duration'' Gamma Ray Bursts (GRBs)  are associated with 
supernova explosions (Bloom et al. 1999; Amati et al. 2000;  Antonelli et al. 2000; 
Piro et al. 2000, Reeves et al. 2002; Hjorth et al. 2003; Price et al. 2003;  
Stanek et al. 2003; Lipkin et al. 2003).         
Particularly, in the case of the gamma ray burst of July 5, 1999 
(GRB990705), in the case of GRB020813 and of  GRB011211, 
it has been possible to estimate the time delay between the two events. 

For GRB990705 Amati et al.  (2000) evaluated that the supernova explosion (SNE) 
has  occurred about 10 years before the GRB,  while Lazzati et al. (2001), 
giving a different interpretation of the same observational data, deduced a time delay 
of about one year.      
In the case of  GRB020813 the supernova event has been estimated 
(Butler et al. 2003)  to have occurred a few months  before the GRB, 
while in the case of  GRB011211 about four days before the burst 
(Reeves et al. 2002).  
If a time-delay between a SNE  and the associated GRB will be    
confirmed by further and more accurate observations, 
thus it is necessary to have a  two-step process.   
The first of these process  is the supernova explosion which forms  
a compact stellar remnant, {\it i.e.} a neutron star. 
The second catastrophic event is associated with the neutron star 
and it is the energy source for the observed GRB.     
These new observational data, and the two-step  scenario outlined above,  
poses severe problems for most of the current theoretical models for  
the central energy source (the so called ``central engine'') of GRBs.  

In a recent paper Berezhiani et al.  (2003) have given a simple and natural 
interpretation of the ``delayed'' Supernova-GRB connection in terms of the 
stellar conversion model (hereafter the {\it Quark Deconfinement Nova} 
 (QDN))  discussed in the present work.      
Here,  with respect to the work of Berezhiani et al. (2003),  we have considered 
two different parameterizations (GM1 and GM3) for the EOS of the hadronic phase,  
and we have explored a larger range for the bag constant in the  EOS 
for the quark phase.   
Moreover, in the present paper the nucleation time has been calculated 
by considering the quantum tunneling of a virtual drop of quark matter in the 
so called Q*-phase (see sect. 2), contrary to the work of Berezhiani et al. (2003) 
where quark flavor conservation during the deconfinement transition has 
been neglected.  We have verified  that flavor conservation in computing the 
nucleation time produces sizable differences in the value  of the critical mass  $M_{cr}$ 
and on the energy released during the QDN which powers the GRB.       

As we can see from the results reported in Tab.s 1 and 2, the total energy ($E_{conv}$)  
liberated during the stellar conversion process is in the range  
0.5--1.7$\times 10^{53}$~erg.    
This huge amount of energy will be mainly carried out by the neutrinos 
produced during the stellar conversion process. 
It has been pointed out by  Salmonson \& Wilson (1999) that near the surface of  
a compact stellar object, due to general relativity effects,  the efficiency 
of the neutrino-antineutrino annihilation into $e^+e^-$ pairs is strongly 
enhanced with respect to the Newtonian case, and it could be as high as 10$\%$.  
The total energy deposited into the electron-photon plasma can therefore be
of the order of $10^{51}$--$10^{52}$ erg.  

The strong magnetic field of the compact star will affect the motion of the 
electrons and positrons, and in turn could generate an  anisotropic  
$\gamma$-ray emission along the stellar magnetic axis.  
%%%  REV.VERSION 01/04 %%%%%%%
This picture is strongly  supported   by  the analysis of the early optical afterglow 
for GRB990123 and GRB021211 (Zhang et al. 2003),   
and by the recent discovery of an ultra-relativistic   
outflow from a  ``neutron star'' in a binary stellar  system (Fender et al. 2004). 
%%%%%%%%%%%%%%%%%%%%%%%%% 
Moreover, it has been recently shown (Lugones et al. 2002)  that the stellar magnetic  
field could influence the velocity of the ``burning front'' of hadronic matter 
into quark matter. This results in a strong geometrical asymmetry of the forming 
quark matter core along the direction of the stellar magnetic axis, 
thus providing a suitable mechanism to produce a collimated GRB 
(Lugones et al. 2002).  Other anisotropies in the GRB could be generated 
by the rotation of the star.

\section{Summary}

In the present work, we have investigated the consequences  of the hadron-quark 
deconfinement phase transition in stellar compact objects when finite size effects 
between the deconfined quark phase and the hadronic phase are taken into account.  
We have found that above a threshold value of the gravitational mass    
a pure hadronic star is metastable to the decay (conversion) to a hybrid neutron star 
or to a strange star
{\footnote{~The particular type of quark star ({it i.e.} hybrid star or strange star) 
formed at the end of the stellar conversion, will depend on the details of the 
quark matter EOS (see sect. 3). Here we want to stress that our scenario 
does not require as a necessary condition the fulfilment of the 
Bodmer-Witten hypothesis on the absolute stability of strange matter and, 
thus the existence of strange stars. The delayed stellar conversion process  
described in this paper takes place also in the case a more ``traditional'' 
hybrid star is formed.}}.   %%%%%%REV.VERSION 01/04 %%%%%%  
We have calculated the {\it mean-life time}  of these metastable 
stellar configurations, the critical mass for the hadronic star sequence, 
and have explored how these quantities depend on the details of the EOS for 
dense matter.  
We have introduced an extension of the concept of limiting mass of  compact stars,   
with respect to the classical one given by Oppenheimer \& Volkov (1939). 
We have demonstrated that, within the astrophysical scenario proposed in the 
present work, the existence of compact stars with ``small'' radii (quark stars)  
does not exclude  the  existence of  compact stars with ``large'' radii (pure hadronic 
stars),  and {\it vice versa}.   

Finally, we have shown that our scenario implies, as a natural consequence 
a two step-process which is able to explain the inferred ``delayed'' connection 
between supernova explosions  and GRBs, giving also the correct energy to power GRBs.   

%%%  REV.VERSION 01/04 %%%%%%%
There are various specific  features and predictions of the present model, 
which we briefly mention in the following. 
The second explosion ({\it Quark Deconfinement Nova})  take place in 
a ``baryon-clean'' enviroment due to the  previous SN explosion.  
Is is possible to have different time delays between the two events   
since the {\it mean-life time} of the metastable hadronic star    
depends  on the value of the stellar central pressure.  
Thus the present model   is able to interpret  a time delay   
of a few years (as observed in GRB990705 (Amati et al. 2000;  Lazzati et al. 2001)),   
of a few months (as in the case of GRB020813 (Butler et al. 2003)),     
of a few days (as deduced for  GRB011211 (Reeves et al. 2002)),    
or the nearly simultaneity of the two events (as in the case of  
SN2003dh and GRB030329 (Hjorth et al. 2003)).          
%%%%%%%%%%%%%%%%%%%%

\acknowledgments

It is a pleasure to acknowledge stimulating discussions with  
David Blaschke, Alessandro Drago, Bennett Link, German Lugones, 
and Sergei B. Popov.


\clearpage

%% the file, uncomment the \plotone, \plottwo, and
%% If you need a layout that cannot be achieved with \plotone or
%% \plottwo, you can invoke the graphicx package directly with the
%% \includegraphics command or use \plotfiddle. For more information,
%% please see the tutorial on "Using Electronic Art with AASTeX" in the
%% documentation section at the AASTeX Web site,
%% http://www.journals.uchicago.edu/AAS/AASTeX.
%%
%% The examples below also include sample markup for submission of
%% supplemental electronic materials. As always, be sure to check
%% the instructions to authors for the journal you are submitting to
%% for specific submissions guidelines as they vary from

%% This example uses \plotone to include an EPS file scaled to
%% 80% of its natural size with \epsscale. Its caption
%% has been written to indicate that additional figure parts will be
%% available in the electronic journal.

\newpage 

%%%%%%%%%%%%%%%%%% TABLES  %%%%%%%%%%%%%%%%%%%%%%%%%%%%%%%

%%%%%%%%%%%%%%%%%% TABLE 1 %%%%%%%%%%%%%%%%%%%%%%%%%%%%%%%
\begin{table*}
\begin{center}
\caption{Critical masses and energy released in the conversion process of an HS into a QS 
for several values of the Bag constant and the surface tension.  
Column labelled $M_{QS,max}$  ($M^b_{QS,max}$) denotes the maximum gravitational 
(baryonic) mass of the final QS sequence. 
The value of the critical gravitational (baryonic) mass of the initial HS is reported on 
column labelled $M_{cr}$ ($M^b_{cr}$) whereas those of the mass of the final QS and the 
energy released in the stellar conversion process 
are shown on columns  lallebed $M_{fin}$ and $E_{conv}$ respectively.   
BH denotes those cases in which due to the conversion the initial HS collapses into a  
black hole. 
Units of B and $\sigma$ are MeV/fm$^3$ and MeV/fm$^2$  respectively. 
All masses are given in solar mass units and the energy released is given in units of  
$10^{51}$ erg. 
The hadronic phase is described with the GM1 model, 
$m_s$ and $\alpha_s$ are always taken equal to $150$ MeV and $0$ respectively. 
The GM1 model preditcs a maximum mass for the pure HS of $1.807$ $M_\odot$.}
\vspace{.4 cm}
\label{tab:tab1}
\begin{tabular}{ccccccccccc}                          \hline\hline

 & & & \multicolumn{4}{c}{$\sigma=10$} & \multicolumn{4}{c}{$\sigma=30$} \\
\cline{5-6} \cline{9-10} \\
\multicolumn{1}{c}{B} & \multicolumn{1}{c}{$M_{QS,max}$} &
\multicolumn{1}{c}{$M^b_{QS,max}$} & \multicolumn{1}{c}{$M_{cr}$} &
\multicolumn{1}{c}{$M^b_{cr}$} &
\multicolumn{1}{c}{$M_{fin}$} & \multicolumn{1}{c}{$E_{conv}$} 
& \multicolumn{1}{c}{$M_{cr}$} &
\multicolumn{1}{c}{$M^b_{cr}$} &
\multicolumn{1}{c}{$M_{fin}$} & \multicolumn{1}{c}{$E_{conv}$} \\
\hline
\\
%%304.88  & 1.807 & 2.053 &  -  &  -  & \multicolumn{2}{c}{-} &   -  &  -  & \multicolumn{2}{c}{-} \\
%%253.11  & 1.807 & 2.053 &  -  &  -  & \multicolumn{2}{c}{-} &   -  &  -  & \multicolumn{2}{c}{-} \\
208.24  & 1.769 & 2.002 &  1.798  &  2.040  & \multicolumn{2}{c}{BH} & 1.805 & 2.050  & 
\multicolumn{2}{c}{BH} \\
169.61  & 1.633 & 1.828 &  1.754  &  1.983  & \multicolumn{2}{c}{BH} & 1.778 & 2.014  & 
\multicolumn{2}{c}{BH} \\
136.63  & 1.415 & 1.561 &  1.668  &  1.871  & \multicolumn{2}{c}{BH} & 1.719 & 1.937  & 
\multicolumn{2}{c}{BH} \\
108.70  & 1.426 & 1.604 &  1.510  &  1.673  & \multicolumn{2}{c}{BH} & 1.615 & 1.804  & 
\multicolumn{2}{c}{BH} \\
106.17  & 1.433 & 1.617 &  1.490  &  1.649  & \multicolumn{2}{c}{BH} & 1.602 & 1.788  & 
\multicolumn{2}{c}{BH} \\
103.68  & 1.441 & 1.632 &  1.469  &  1.623  & 1.434 & 62.5 & 1.588 & 1.770  & 
\multicolumn{2}{c}{BH} \\
101.23  & 1.449 & 1.648 &  1.447  &  1.596  & 1.411 & 64.0 & 1.574 & 1.752  & 
\multicolumn{2}{c}{BH} \\
98.83  & 1.459 & 1.665 &  1.425  &  1.569  & 1.388 & 66.0 & 1.559 & 1.734  & 
\multicolumn{2}{c}{BH} \\
96.47  & 1.470 & 1.684 &  1.402  &  1.541  & 1.364 & 68.5 & 1.543 & 1.715  & 
\multicolumn{2}{c}{BH} \\
94.15  & 1.481 & 1.704 &  1.378  &  1.513  & 1.339 & 71.1 & 1.527 & 1.694  & 
1.474 & 94.8 \\
91.87  & 1.494 & 1.726 &  1.354  &  1.484  & 1.313 & 74.2 & 1.511 & 1.674  & 
1.456 & 98.1 \\
89.64  & 1.507 & 1.750 &  1.329  &  1.453  & 1.285 & 77.3 & 1.495 & 1.654  & 
1.438 & 101.8 \\
87.45  & 1.552 & 1.776 &  1.302  &  1.422  & 1.257 & 80.7 & 1.477 & 1.632  & 
1.417 & 105.9 \\
85.29  & 1.538 & 1.803 &  1.275  &  1.389  & 1.228 & 84.4 & 1.458 & 1.610  & 
1.397 & 110.4 \\
80.09  & 1.581 & 1.879 &  1.196  &  1.296  & 1.144 & 92.9 & 1.410 & 1.551  & 
1.342 & 122.7 \\
75.12 & 1.631 & 1.966 &  1.082  &  1.164  & 1.029 & 93.78 & 1.359 & 1.489  & 
1.284 & 133.1 \\
65.89  & 1.734 & 2.156 &  0.820  &  0.867  & 0.764 & 100.6 & 1.212 & 1.315  & 
1.123 & 159.9 \\
63.12  & 1.770 & 2.222 &  0.727  &  0.764  & 0.672 & 98.1 & 1.160 & 1.254  & 
1.067 & 166.5 \\
59.95  & 1.814 & 2.305 &  0.545  &  0.566  & 0.501 & 79.7 & 1.081 & 1.162  & 
0.986 & 168.8 \\

\\
\hline
\end{tabular}
\end{center}
\end{table*}

%%%%%%%%%%%%%%%%%% TABLE 2 %%%%%%%%%%%%%%%%%%%%%%%%%%%
\begin{table*}
\begin{center}
\caption{Same as Table 1 for the GM3+Bag model. 
The situations  for which there is no deconfinement phase transition,  or 
for wich the nucleation time of the hadronic maximum mass configuration is of the order 
or larger that the age of the universe (see discusion in the text) are reported with no entry ( - )
The maximum mass for the pure HS predicted by the GM3 model is 
$1.552$ $M_\odot$.}
\vspace{.4 cm}
\label{tab:tab2}
\begin{tabular}{ccccccccccc}                          \hline\hline

 & & & \multicolumn{4}{c}{$\sigma=10$} & \multicolumn{4}{c}{$\sigma=30$} \\
\cline{5-6} \cline{9-10} \\
\multicolumn{1}{c}{B} & \multicolumn{1}{c}{$M_{QS,max}$} &
\multicolumn{1}{c}{$M^b_{QS,max}$} & \multicolumn{1}{c}{$M_{cr}$} &
\multicolumn{1}{c}{$M^b_{cr}$} &
\multicolumn{1}{c}{$M_{fin}$} & \multicolumn{1}{c}{$E_{conv}$}   
& \multicolumn{1}{c}{$M_{cr}$} &
\multicolumn{1}{c}{$M^b_{cr}$} &
\multicolumn{1}{c}{$M_{fin}$} & \multicolumn{1}{c}{$E_{conv}$} \\
\hline
\\
136.63  & 1.448 & 1.613 &  -  &  -  & \multicolumn{2}{c}{-} &  1.551   &  1.734  & \multicolumn{2}{c}{BH} \\
122.07  & 1.430 & 1.600 &  -  &  -  & \multicolumn{2}{c}{-} &  1.539   &  1.718  & \multicolumn{2}{c}{BH} \\
108.70  & 1.440 & 1.627 &  1.484  &  1.648  & \multicolumn{2}{c}{BH} &  1.498   &  1.665  & \multicolumn{2}{c}{BH} \\
106.17  & 1.445 & 1.637 &  1.474  &  1.635  & 1.444 & 53.4 & 1.487  &  1.651  & \multicolumn{2}{c}{BH} \\
103.68  & 1.451 & 1.648 &  1.461  &  1.619  & 1.430 & 56.2 & 1.475  &  1.636  & 1.442 & 57.9 \\
101.23  & 1.458 & 1.661 &  1.449  &  1.603  & 1.415 & 59.3 & 1.462  &  1.620  & 1.428 & 60.9 \\
98.83   & 1.466 & 1.676 &  1.435  &  1.587  & 1.400 & 62.5 & 1.449  &  1.604  & 1.413 & 64.3 \\
96.47   & 1.475 & 1.692 &  1.421  &  1.569  & 1.384 & 66.3 & 1.436  &  1.587  & 1.398 & 68.0 \\
94.15   & 1.485 & 1.710 &  1.406  &  1.550  & 1.367 & 70.0 & 1.422  &  1.570  & 1.382 & 71.9 \\
91.87   & 1.497 & 1.731 &  1.390  &  1.531  & 1.348 & 74.2 & 1.407  &  1.552  & 1.365 & 76.2 \\
89.64   & 1.509 & 1.753 &  1.373  &  1.510  & 1.329 & 78.5 & 1.392  &  1.534  & 1.347 & 80.8 \\
87.45   & 1.523 & 1.778 &  1.355  &  1.488  & 1.308 & 83.1 & 1.376  &  1.515  & 1.329 & 85.7 \\
85.29   & 1.538 & 1.804 &  1.335  &  1.463  & 1.285 & 87.8 & 1.360  &  1.495  & 1.310 & 90.6 \\
80.09   & 1.582 & 1.881 &  1.270  &  1.385  & 1.214 & 98.7 & 1.314  &  1.438  & 1.255 & 104.0 \\
75.12   & 1.631 & 1.966 &  1.181  &  1.280  & 1.121 & 107.8 & 1.252  &  1.365  & 1.187 & 116.85 \\
65.89   & 1.734 & 2.155 &  0.943  &  1.005  & 0.877 & 117.8 & 1.126  &  1.216  & 1.045 & 144.5 \\
63.12   & 1.770 & 2.222 &  0.808  &  0.854  & 0.747 & 110.2 & 1.082  &  1.164  & 0.997 & 152.1 \\
59.95   & 1.814 & 2.305 &  0.513  &  0.531  & 0.471 & 74.7 & 1.010  &  1.081  & 0.923 & 155.5 \\

\\
\hline
\end{tabular}
\end{center}
\end{table*}
%%%%%%%%%%%%%%%%%%%%%%%%%%%%%%%%%%%%%%%%%%%%%%%%%%%%%%%

%%%%%%%%%%   TABLE 3   %%%%%%%%%%%%%%%%%%%%%%%
\begin{table*}
\begin{center}
\caption{Critical masses and energy released in the conversion process of an HS into a QS 
for several values of the  curvature coefficient $\gamma$.   
The results are realtive to the GM1 model for the  hadronic phase. 
For the quark phase we take $B = 75.12$~MeV/fm$^3$,  
$m_s = 150$ MeV  and $\alpha_s =  0$.   
The surface tension is   $\sigma = 30$~MeV/fm$^2$.      
All the quantities reported in the table have the same units as in Tab. 1. }   
\vspace{.4 cm}
\label{tab:tab3}
\begin{tabular}{ccccc}                          \hline\hline

\multicolumn{1}{c}{$\gamma$}   & 
\multicolumn{1}{c}{$M_{cr}$}       &
\multicolumn{1}{c}{$M^b_{cr}$}   &
\multicolumn{1}{c}{$M_{fin}$}     & 
\multicolumn{1}{c}{$E_{conv}$} \\
\hline
\\
       0 &  1.359  &  1.489  &   1.284 &   133.1  \\
     10 &  1.390  &  1.526  &   1.313 &   137.9  \\
     20 &  1.418  &  1.561  &   1.338 &   142.3  \\

\\
\hline
\end{tabular}
\end{center}
\end{table*}

%%%%%%%%%%%%   FIGURES    %%%%%%%%%%%%%%%%%%%%%%%%%%%%

%%%%%%%%%%%%   FIGURE 1    %%%%%%%
\clearpage

%%%%%%%%%%   FIGURE 2    %%%%%%%%%%%%
\clearpage  

%%%%%%%%         FIGURE 3      %%%%%%%%%%%%%%%%%%%%%%%%%%%
\clearpage  

%%%%%%%%%%%    FIGURE 4    %%%%%%%%%%%%%%%%%%%%%%%%%
\clearpage 

%%%%%%%%%   FIGURE 5     %%%%%%%%%%%%%%%%%%%%%%%%%%%%%%%%
\clearpage

%%%%%%%%   FIGURE 6   %%%%%%%%%%%%%%%%%%%%%%%
\clearpage

%%%%%%%%   FIGURE 7    %%%%%%%%%%%%%%%%%%%%%%%
\clearpage

%%%%%%%%%     FIGURE 8    %%%%%%%%%%%%%%%%%%%%%%%%
\clearpage

%%%%%%%%%%%%%%%%%% FIGURE 9 %%%%%%%%%%%%%%%%%%%%%%%%%%%%%
\clearpage

%%%%%%%%%%%%%%%%%% FIGURE 10 %%%%%%%%%%%%%%%%%%%%%%%%%%%%%
\clearpage

%%%%%%%%%%%%%%%

\begin{thebibliography}{}

\bibitem{afo86}    Alcock, C.,  Farhi, E., \&  Olinto, A.  1986, \apj, 310, 261

\bibitem{amati}   Amati, L.  et al., 2000, Science, 290, 953 

\bibitem{antonelli}   Antonelli, L.A.  et al.  2000,  ApJ, 545,   L39  

\bibitem{bc76}  Baym, G., \&  Chin, S.A. 1976, Phys. Lett. B, 62, 241 

\bibitem{bh89}   Benvenuto, O.G.,  \& Horvath, J.E.  
                                 1998, Phys. Rev. Lett. , 63, 716   %%% 

\bibitem{QDN}  Berezhiani, Z., Bombaci, I., Drago, A., Frontera, F., \& Lavagno, A. 
                               2003, ApJ, 586, 1250  

\bibitem{bign03}   Bignami, G.F.,  Carvero, P.A.,   De Luca,  A., \&   Mereghetti, S., 
                                    2003, Nature,  423,   725  

\bibitem{bloom99}   Bloom, J.S.  et al.  1999, Nature, 401, 453 

\bibitem{bod71} Bodmer, A.R.  1971, Phys. Rev. D, 4, 1601

\bibitem{bom96} Bombaci, I., 1996, A \&A , 305, 871

\bibitem{bom97} Bombaci, I., 1997, Phys. Rev. C, 55, 1587

\bibitem{bom03} Bombaci, I., 2003, astro-ph/0307522

\bibitem{bom04} Bombaci, I., et al. 2004,  work in progress 

\bibitem{bd00}   Bombaci, I., \& Datta, B. 2000, ApJ, 530, L69 

\bibitem{but03}  Butler, N.R., et al.  2003, ApJ, 597, 1010  %%%%

\bibitem{c98} Cheng, K.S., Dai, Z.G., Wai, D.M., \& Lu, T., 1998, Science, 280, 407

\bibitem{cp75}  Collins,  J.C., \&  Perry,  M.J.  1975,  Phys. Rev. Lett.,  34, 1353 

\bibitem{cott02}   Cottam, J., Paerels, F., \& Mendez, M. 2002, Nature, 420, 51

\bibitem{d98} Dey, M., Bombaci, I.,  Dey, J., Ray, S., \& Samanta, B.C. 1998, 
                                   Phys. Lett. B, 438, 123; erratum 1999, Phys. Lett. B, 467, 303

\bibitem{DL01}   Drago, A., \& Lavagno, A.  2001, Phys. Lett. B, 511, 229

\bibitem{drake02}  Drake, J.J.,  et al. 2002,  ApJ, 572, 996

\bibitem{fj84} Farhi, E., \& Jaffe, R.L. 1984, Phys. Rev. D, 30, 2379  

\bibitem{fend04}    Fender, R., et al. 2004, Nature, 427, 222  %%%%

\bibitem{grassi}    Grassi, F. 1998,   ApJ, 492, 263 %%%%

\bibitem{gle92} Glendenning, N.K 1992, Phys. Rev. D, 46, 1274

\bibitem{gle96}  Glendenning,  N.K., 1996,  Compact Stars: Nuclear Physics, 
                               Particle Physics, and General Relativity, Springer Verlag 

\bibitem{gm91} Glendenning, N.K., \& Moszkowski, S.A. 1991, Phys. Rev. Lett., 67, 2414 

\bibitem{hae92} Haensel, P., Zdunik, J.L. \& Schaefer, R. 1986, A\&A, 160, 121

\bibitem{hae03} Haensel, P. 2003,  Equation of state of dense matter and 
                                maximum mass of neutron stars,  
                                in Final Stages of Stellar Evolution, Ed. C. Motch and  J.-M. Hameury,  
                                 EAS Publications Series 7,  249  

\bibitem{hei93}   Heiselberg, H., Pethick, C.J., \& Staubo, E.F. 1993, 
                                Phys. Rev. Lett., 70, 1355

\bibitem{hei95}   Heiselberg, H.  1995,  in Strangeness and Quark Matter, 
                                 World Scientific,  338  %%%%%

\bibitem{hjo03}    Hjorth, J. et al,   2003, Nature, 423,  847  

\bibitem{hbv92}   Horvath, J.E.,   Benvenuto, O.G.,  \& Vucetich, H.  
                                 1992, Phys. Rev. D, 45, 3865    %%% 

\bibitem{horv94}  Horvath,  J.E.   1994, Phys. Rev. D, 49, 5590   %%% 

\bibitem{iach}  Iachello, F., Langer, W.D.,  \& Lande, A. 1974, Nucl. Phys. A, 219, 612 

\bibitem{is97} Iida, K., \& Sato, K. 1997, Prog. Theor. Phys., 1, 277: 
                                                                            1998, Phys. Rev. C, 58, 2538
\bibitem{itoh}  Itoh, N. 1970, Prog. Theor. Phys., 44, 291 

\bibitem{ik69}  Ivanenko, D., \&  Kurdgelaidze, D.F. 1969, Lett. Nuovo Cimento, 2, 13 

\bibitem{kapl02}   Kaplan, D.L., van Kerkwijk, M.H, \&  Anderson, J.   2002, ApJ, 571, 447

\bibitem{kk76}  Keister, B.D., \&  Kisslinger, L.S., 1976, Phys. Lett. B, 64, 117 

\bibitem{lazz01}  Lazzati, D., Ghisellini, G., Amati, L., Frontera, F., Vietri, M., \& Stella, L. 
                                 2001, ApJ, 556, 471  

\bibitem{li99a}     Li, X.--D., Bombaci, I., Dey, M., Dey J., \& van den Heuvel, E.P.J. 
                                   1999a, Phys. Rev. Lett., 83, 3776  

\bibitem{li99b}   Li, X.--D., Ray, S., Dey, J., Dey, M., \& Bombaci, I. 1999b, 
                               \apj, 527, L51 

\bibitem{lk72}     Lifshitz, I. M., \& Kagan, Y. 1972, Sov. Phys. JETP, 35, 206  

\bibitem{lip03}    Lipkin, Y.M.   2003,  astro-ph/0312594

\bibitem{lb98}     Lugones, G.,  \&   Benvenuto, O.G. 1998 , Phys. Rev. D, 58, 083001  %%%   

\bibitem{lug02}   Lugones, G., Ghezzi,  C.R.,  de Gouveia Dal Pino E.M., \&  
                                 Horvath, J.E. 2002, ApJ, 581, L101  

\bibitem{mad93}   Madsen, J. 1993,  Phys. Rev. Lett., 70, 391  

\bibitem{madsen}   Madsen, J. 1999,  Lectures Notes in Physics Vol. 500, 
                                      Springer Verlag, 162

\bibitem{muller}   M\"{u}ller, H., \& Serot, B.D. 1995, Phys. Rev. C, 52, 2072

\bibitem{om94}     Olesen, M.L.,  \&  Madsen, J.   1994, Phys. Rev. D, 49, 2698   %%% 

\bibitem{ov39}    Oppenheimer,  J.R., \& Volkoff, G.M. 1939, Phys. Rev., 55, 374  

\bibitem{ok99}   Orosz, J.A., \& Kuulkers, E.  1999, MNRAS, 305, 132 

\bibitem{oz03}  \"{O}zel, F.  \&  Psaltis, D. 2003,  ApJ, 582, L31

\bibitem{piro}    Piro, L.  et al.    2000, Science,  290,  955

\bibitem{pout03}    Poutanen, J.  \& Gierli\`nski, M.  2003,  MNRAS, 343, 1301 

\bibitem{prak97}   Prakash, M., Bombaci, I., Prakash, M., Ellis, P.J.,  Lattimer, J.M., 
                                    \&   Knorren, R. 1997,  Phys. Rep. 280, 1  %%%
 
\bibitem{price03}    Price, P.A. et al.  2003,  Nature, 423, 844 

\bibitem{quai03} Quaintrell, H., et al.,  2003,  A\&A, 401, 313 

\bibitem{reeves02}   Reeves, J.N.  et al.,  2002, Nature,  414,  512

\bibitem{roger88}  Roget, R.S., Milne, D.K., Kesteven, M.J., Wellington, K.J., \& 
                                    Haynes, R.F. 1988, ApJ, 332, 940  

\bibitem{salm99}   Salmonson, J.D.,  \&  Wilson, J.R.  1999,  ApJ,  517,  859 

\bibitem{sanw02}   Sanwal, D., Pavlov, G.G., Zavlin, V.E., \& Teter, M.A. 2002, ApJ, 574, L61 

\bibitem{st83}  Shapiro,  S.L. \& Teukolsky, S.A. 1983, 
                                          Black holes, white dwarfs and neutron stars, 
                                          Ed.  J. Wiley\& Sons 

\bibitem{ST02} Shaposhnikov, N. , \&  Titarchuk, L.  2002, ApJ, 570, L25

\bibitem{shapo} Shaposhnikov, N. ,  Titarchuk, L., \& Haberl, F. 2003, 
                                              ApJ, 593, L38 (STH)  

\bibitem{spe03}      Spergel, D.N. 2003,  Astrophys. J. Suppl.,  148, 175

\bibitem{stan03}   Stanek, K.Z. et al. 2003, ApJ,  591, L17 

\bibitem[1979]{ter79} Terazawa, H. 1979, INS-Report, 336 (INS, Univ. of Tokyo); 
                                          1989, J. Phys. Soc. Japan, 58, 3555; 
                                         1989, J. Phys. Soc. Japan, 58, 4388

\bibitem{tc99} Thorsett, S.E., \& Chakrabarty D.  1999, ApJ, 512, 288 

\bibitem[1994]{titarchuk94} Titarchuk, L.  1994,  ApJ,  429, 330 

\bibitem{vkerk95} van Kerkwijk, M.H. \etal 1995, A\&A 303, 483  

\bibitem{vyt03}    Voskresensky, D.N., Yasuhira, M, \& Tatsumi, T.  2003, 
                                   Nucl. Phys. A, 723, 291

\bibitem{wl02}   Walter, F.M., \& Lattimer, J.M. 2002, ApJ, 576, L148  

\bibitem{witt84} Witten, E. 1984, Phys. Rev. D, 30, 272

\bibitem[2002]{xu02}  Xu, R.X.  2002, ApJ, 570, L65

\bibitem[2003]{xu03}  Xu, R.X.  2003,  Chin. J. Astron. Astrophys., 3, 33 


\bibitem{zhang03}   Zhang, B., Kobayashi, S., \& Meszaros, P.  2003, ApJ, 595, 950 %%%



\end{thebibliography}
\end{document}